\documentclass[acmsmall, nonacm]{acmart}

\usepackage{graphicx}
\usepackage{pdfpages}
\usepackage[title]{appendix}

\usepackage{textcomp}
\usepackage{xcolor}
\def\BibTeX{{\rm B\kern-.05em{\sc i\kern-.025em b}\kern-.08em
    T\kern-.1667em\lower.7ex\hbox{E}\kern-.125emX}}
    
\usepackage{multicol}
\usepackage[utf8]{inputenc} 
\usepackage{colortbl}
\graphicspath{{Figures/}}
\usepackage{afterpage}
\usepackage{lscape}
\usepackage{multirow}
\usepackage{soul,color}
\usepackage{verbatim}
\usepackage{array,multirow,makecell}
\usepackage[hyphenbreaks]{breakurl}

\usepackage{adjustbox}
\usepackage{tikz}
\usepackage{pgfplots}
\usepackage{pgfplotstable}
\usetikzlibrary{plotmarks}
\usetikzlibrary{shapes.multipart}
\usetikzlibrary{patterns}
\usetikzlibrary{fadings}
\usetikzlibrary{shapes.callouts}
\usepackage{pgfkeys}
\usetikzlibrary{decorations.pathmorphing}
\usetikzlibrary{decorations.markings}
\usetikzlibrary{shadows}
\usetikzlibrary{shapes}
\usepackage{hyperref}
\usepackage{subcaption}
\usepackage{pdflscape}

\usepackage{listings}
\usepackage{color}
\IfFileExists{./useminted}{\usepackage[frozencache=true,cachedir=mintedpapier]{minted}}{}

\definecolor{dkgreen}{rgb}{0,0.6,0}
\definecolor{gray}{rgb}{0.5,0.5,0.5}
\definecolor{mauve}{rgb}{0.58,0,0.82}

\usepackage{adjustbox}
\usepackage{longtable,tabularx,ltxtable,ragged2e}
\usepackage{filecontents}
\usepackage{color}
\usepackage{ifthen}
\usepackage{soulutf8} 
\usepackage{numprint}
\usepackage{pgf-pie}

\newcommand{\numberOfJVMs}{\numprint{147}}
\newcommand{\numberOfExperiments}{\numprint{256515}}
\newcommand{\totalNumberOfLibrariesVersions}{\numprint{1,410}}
\newcommand{\totalNumberOfYsoserialPayloads}{\numprint{34}}
\newcommand{\NumberOfStudiedLibraries}{\numprint{14}}
\newcommand{\MitreCVEs}{\numprint{361}}
\newcommand{\TotalCVEs}{\numprint{364}}
\newcommand{\JavaDeserCVEs}{\numprint{104}}
\newcommand{\NumberStudiedAttacks}{\numprint{19}}
\newcommand{\DVCVEs}{\numprint{95}}
\newcommand{\NBApps}{\numprint{77}}
\newcommand{\SuccessfulCVES}{\numprint{58}}

\newcommand{\myconclusion}[1]{                                                  
\begin{center}
\begin{tikzpicture}
\node[draw, rounded corners, fill=black!10, inner sep=.25cm, outer sep=0cm] {
\begin{minipage}{0.9\columnwidth}
#1
\end{minipage}
};
\end{tikzpicture}
\end{center}
}

\let\llncssubparagraph\subparagraph
\let\subparagraph\paragraph
\usepackage[compact]{titlesec}
\let\subparagraph\llncssubparagraph

\titlespacing\section{0pt}{4pt plus 4pt minus 2pt}{4pt plus 2pt minus 2pt}
\titlespacing\subsection{0pt}{4pt plus 4pt minus 2pt}{4pt plus 2pt minus 2pt}
\titlespacing\subsubsection{0pt}{4pt plus 4pt minus 2pt}{4pt plus 2pt minus 2pt}

\newenvironment{myminted}[2][def]
{\IfFileExists{./useminted}{\minted[#1]{#2}}{\scriptsize\verbatim}}
{\IfFileExists{./useminted}{\endminted}{\endverbatim}}

\makeatletter
\let\@authorsaddresses\@empty
\makeatother

\begin{document}

\title{An In-depth Study of Java Deserialization Remote-Code Execution  Exploits and Vulnerabilities}

\author[Imen Sayar]{Imen Sayar$^\dagger$}
\email{imen.sayar@irit.fr}
\affiliation{
  \institution{University of Toulouse} 
  \city{Blagnac} 
  \country{France}
  \postcode{31070}
}
\thanks{$^{\dagger}$Part of this research was conducted when Imen Sayar was at the University of Luxembourg}

\author[Alexandre Bartel]{Alexandre Bartel$^*$}
\email{alexandre.bartel@cs.umu.se}
\affiliation{
  \institution{Umeå University}
  \streetaddress{MIT-Huset}
  \city{Umeå}  
  \country{Sweden} 
}
\thanks{$^{*}$Part of this research was conducted when Alexandre Bartel was at the University of Luxembourg and the University of Copenhagen.}

\author{Eric Bodden}
\email{eric.bodden@uni-paderborn.de}
\affiliation{
  \institution{Paderborn University}
  \city{Paderborn}  
  \country{Germany} 
}

\author{Yves Le Traon}
\email{Yves.LeTraon@uni.lu}
\affiliation{
  \institution{University of Luxembourg}
  \streetaddress{6, rue Richard Coudenhove-Kalergi}
  \city{Kirchberg Campus} 
  \country{Luxembourg}
  \postcode{L-1359}
}

\begin{abstract}

Nowadays, an increasing number of applications uses deserialization. This technique, based on rebuilding the instance of objects from serialized byte streams, can be dangerous since it can open the application to attacks such as remote code execution (RCE) if the data to deserialize is originating from an untrusted source. Deserialization vulnerabilities are so critical that they are in OWASP's list of top 10 security risks for web applications. This is mainly caused by faults in the development process of applications and by flaws in their dependencies, i.e., flaws in the libraries used by these applications. No previous work has studied deserialization attacks in-depth: How are they performed? How are weaknesses introduced and patched?  And for how long are vulnerabilities present in the codebase?
To yield a deeper understanding of this important kind of vulnerability, we perform two main analyses: one on attack gadgets, i.e., exploitable pieces of code, present in Java libraries, and one on vulnerabilities present in Java applications. 
For the first analysis, we conduct an exploratory large-scale study by running \numberOfExperiments ~experiments in which we vary the versions of libraries for each of the \NumberStudiedAttacks{} publicly available exploits.
Such attacks rely on a combination of \emph{gadgets} present in one or multiple Java libraries. A gadget is a method which is using objects or fields that can be attacker-controlled.
Our goal is to precisely identify library versions containing gadgets and to understand how gadgets have been introduced and how they have been patched.
We observe that the modification of one innocent-looking detail
in a class -- such as making it \texttt{public} -- can already introduce a gadget. Furthermore, we noticed that among the studied libraries, 37.5\%  are not patched, leaving gadgets available for future attacks. 

For the second analysis, we manually analyze 104 deserialization vulnerabilities CVEs to understand how
vulnerabilities are introduced and patched in real-life Java applications.
Results indicate that the vulnerabilities are not always completely patched or
that a workaround solution is proposed.
With a workaround solution, applications are still vulnerable since the code itself is unchanged.

\end{abstract}


\maketitle

%-------------------------------------------------------------------------------
\section{Introduction}
\label{sec:introduction}

Over the past 10 years, the MITRE Corporation~\cite{mitreDeserialization} 
registered \textit{364} CVEs linked to deserialization vulnerabilities\footnote{All the queries have been done in June 2021}
in several mainstream programming languages such as Java, PHP, and .NET.
These critical vulnerabilities, frequently allowing Remote Code Execution  (RCE), are a highly relevant topic in the research community.
For instance, Shcherbakov \textit{et al.} \cite{SerialDetector} recently developed an open-source tool named \textit{SerialDetector} allowing the detection of deserialization vulnerabilities in .NET applications.
In this paper, we focus on the characterization of Java deserialization vulnerabilities. These vulnerabilities result because of flaws existing in the   applications' development process or in the libraries used by these applications. 

Java serialization allows  transforming class instances into a stream of bytes. 
Java objects can therefore be transferred through a network.  
Deserialization  consists of reading the serialized byte stream in order to rebuild the original instances by loading their fields. 
While serialization is convenient to transfer objects between hosts, 
it presents a danger when the source of the data to deserialize is untrusted.
Indeed, an attacker could craft a byte stream that, when deserialized on the remote host,
could control the execution flow of the Java code by chaining sequences of Java code called \emph{gadgets}.
A deserialization attack can be performed by leveraging either gadgets present in the Java Class Library (JCL)
or in an external library 
or by combining many libraries containing gadgets, that we will refer to as \emph{gadget libraries} in the remainder of this paper.

Figure~\ref{fig:simple-example} shows a simple example of a Java serializable  class \texttt{A} characterized by a \texttt{String} field called \texttt{command}. When calling the \texttt{writeObject} method, suppose that the attacker changes this field with the Linux command line \texttt{top} (line number 6). The execution of \texttt{readObject} leads to the call of the gadget \texttt{Runtime.exec} (line 9)  which is using the attacker-controlled object \texttt{command}.

\begin{figure}[ht]
\centering
\begin{myminted}[linenos=true,xleftmargin=15pt, fontsize=\scriptsize]{Java}
import java.io.Serializable;
import java.lang.Runtime;
public class A implements Serializable {
	String command; 
	private void writeObject() {
		command = "top"; //the attacker can change this field
	}
	private void readObject() {
		Runtime.exec(command); //the attacker can execute the command
	}
}
\end{myminted}
\caption{A simple example of a serializable class. 
In this class, the attacker can modify the  command field. 
Thus, during deserialization, when the JVM calls A.readObject() method, the attacker command will be executed instead of ``top''.} 
\label{fig:simple-example}
\end{figure}

In 2015, Frohoff and Lawrence demonstrated 
how to exploit unsafe Java deserialization vulnerabilities \cite{marshalling-pickles}. 
The same year, Litchfield~\cite{paypalBug} and Stepankin~\cite{stepankin}
identified an RCE Java deserialization vulnerability in one of PayPal’s critical applications, 
the manager portal\footnote{\url{manager.paypal.com}} which could allow attackers to reach 
production databases.
In 2016, an attacker took control of \numprint{2000} computers of the 
Metropolitan Transport Agency of San Francisco
through a Java deserialization vulnerability in the Web server~\cite{sfmta}. 
Equifax\footnote{\url{https://www.equifax.com/}} had one of its worst bugs in 2017 when attackers infiltrated its network and stole the personal information of  147.7 million Americans from its servers. 
The entry point of this attack was \emph{CVE-2017-9805}, 
a Java deserialization vulnerability in Apache Struts' 
web application~\cite{equifaxAttack}.  
All these concrete real-world examples support the conclusion of multiple studies~\cite{OWASP,hdivsecurity} 
ranking insecure deserialization in the top 10 of the most dangerous web application security vulnerabilities. More precisely, in 2021, OWASP classifies this kind of vulnerability in the 8th position after other dangerous vulnerabilities like Cross-Site Scripting (XSS)~\cite{OWASP2021} -classified as the third most dangerous vulnerability-  or buffer overflow~\cite{OWASPBufferOverflow}. We study this problem of deserialization attacks because - compared to other vulnerabilities-  a deserialization attack is able to completely control the victim systems or to give place to ransomware attacks. For instance, on one hand, a buffer overflow on modern operating systems will not give the attacker anything on its own because it needs to be chained with at least an information leak and other vulnerabilities to bypass other mitigation techniques.
On the other hand, a deserialization vulnerability is at a higher level and might allow the attacker a complete control over the target system.  Another important aspect is that "dormant" serialization vulnerabilities can be super easily enabled, once new gadgets are accidentally introduced. To our point of view, it is much more subtle than  other kinds of vulnerabilities such as injections.

In this paper, we study Java gadgets and Java deserialization vulnerabilities found in real-world applications 
leveraging the standard Java deserialization mechanism~{\cite{riggs1996pickling}}.
The first study is based on the analysis of gadgets from 19 publicly available remote code execution (RCE) attacks from the ysoserial Github repository~\cite{ysoserial}. Ysoserial is a project that gives a proof of concept tool and provides  34 Java  payloads exploited in publicly known deserialization attacks. These latter are carried out  by chaining gadgets. In this paper, we focus on 19 RCE attacks representing the majority of the ysoserial attacks.  Our study is limited to this kind of attacks because we have developed a framework detecting RCE attacks only. 
The second study is based on the manual analysis of \NBApps ~Java applications impacted by a Java deserialization vulnerability described in a CVE~\footnote{CVE refers to Commons Vulnerabilities Exposures. According to the Mitre terminology, a CVE is identified using an ID which is "a unique, alphanumeric identifier assigned by the CVE Program. Each identifier references a specific vulnerability."~\citep{oldMitre,newBetaMitre}}.  
We found that for all attacks relying on a \emph{single} gadget library, this
library has been patched. The patching action can impact 
 one or many gadget libraries involved in an attack, i.e.,  when the attack relies on multiple  gadget libraries, patching a single library may be sufficient to avoid \emph{this} attack. 
Yet, the non-patched gadget libraries can often still be leveraged later if they can be combined with other gadgets. Thus, even though they might not lead to an exploitable software at the time, they do increase the attackers' capabilities, thus weakening the software system. One aspect of our analysis is the  detection of recent gadget library versions. This is relevant since it points to  recent library versions non-cited in ysoserial repository, like the commons-beanutils,  that still contain gadgets.
This may alert developers to be aware of these gadgets' library versions if used in the classpath of their applications and to check if the concerned library versions are mentioned as gadgets ones in CVEs database.

Furthermore, this paper analyzes how gadgets are introduced into libraries---an important point that previous pieces of research did not explicitly address.  When analyzing the 19 RCE exploits, we have identified several ways to introduce a gadget in a library:  
adding classes, methods, and interfaces, or changing the signature of methods. 
Our main conclusion is that the modification of one innocent-looking detail in a class -- such as making it \texttt{public} --  can already introduce a gadget.
When studying patches of such libraries, we observed that the time used to remove gadgets varies between several months and almost 12 years, with an average of almost six years. It thus appears that deserialization vulnerabilities do not yet get the attention of practitioners that they should actually deserve.

The study on Java applications clearly shows that developers should never write code that deserializes data from an untrusted source because it becomes an obvious entry point for attackers.
Solutions exist to prevent knows attacks, e.g. allow/deny lists, but they are not fool-proof, as the complete list of gadgets present in Java libraries is unknown.
Interestingly enough, for 24.1\% of the studied CVEs, the solution that has been selected to prevent the exploitation of the vulnerability is \emph{not} a code change but a workaround. Workarounds work well on an already deployed system, however, they might not be applied in new deployments or in a new software environment, which makes the vulnerability accessible to the attacker again.

This paper is concerned with how deserialization vulnerabilities in Java manifest in practice. We present the following contributions: 
\begin{itemize}
\item We conduct a large-scale study on more than \numberOfExperiments~ combinations of  \NumberOfStudiedLibraries ~libraries, representing 
\NumberStudiedAttacks{} publicly known Java deserialization RCE exploits, and \numberOfJVMs ~Java runtimes to understand which precise library versions introduce gadgets, how they are patched, and the structure of attacks in terms of gadgets. A thorough description of the experimental procedure used to obtain the experimentation data, including how test subjects were collected, is also described.

 \item We detail how deserialization gadgets  are introduced in the libraries. To our knowledge, this is the first work to consider how deserialization vulnerabilities manifest in real code bases and libraries. This provides some insight into how these vulnerabilities are commonly treated outside of academia.
 
 \item Based on the results of our analysis on how gadgets are introduced, we propose recommendations for library developers to prevent the introduction of gadgets. 
 
 \item We perform  a study of the patching time of some libraries and show that it can sometimes take over 10 years.

\item We perform an analysis of 104 Mitre CVEs that concern  deserialization vulnerabilities in Java applications and conclude that not all the patches  prevent the attacks and protect the applications. 
 
\end{itemize}

The remainder of this paper is organized as follows. Section~\ref{sec:insecure-deserialization} presents  background about essential concepts related to Java deserialization vulnerabilities and uses an example to explain how a deserialization attack can be performed. 
 Section~\ref{sec:methodandeval} explains our methodology  and details our two analyses: 
 the large-scale study about attack gadgets present in Java libraries and 
 the analysis of vulnerabilities present in Java applications with regard  to the libraries and JVM versions.  Take-away messages and lessons learned from our analyses are described in Section~\ref{sec:synthesis}. Section~\ref{sec:limitations} points on the limitations of our approach.  The state of the art is discussed in Section~\ref{sec:state-art}. Finally, Section~\ref{sec:conclusion} concludes this work.

%-------------------------------------------------------------------------------

%-------------------------------------------------------------------------------
\section{Background}  
\label{sec:insecure-deserialization}
 \subsection{Terminology}
 \label{ser-deser-mechanisms}
 
Before starting the study of Java deserialization attacks, we define the terminology used all along this paper. 
 
 \paragraph{\textbf{Vulnerability.}} 
We use Mitre's definition~\cite{MitreVulnerability}: 
\emph{"[A vulnerability is] a flaw in a software, firmware, hardware, or service component resulting from a weakness that can be exploited, causing a negative impact to the confidentiality, integrity, or availability of an impacted component or components."}.

 \paragraph{\textbf{Gadget and gadget chain.}} 

In the context of this paper, a gadget is a Java method using objects or fields that can be attacker-controlled. A gadget chain is a malicious sequence of method (gadget) calls created by an attacker. The presence of a set of gadgets in the classpath of a vulnerable application is one of the conditions required to carry out deserialization attacks.   
 
\paragraph{\textbf{Deserialization vulnerability in Java.}} 
A Java deserialization vulnerability is a weakness in the code that can be exploited when the Java code deserializes an attacker-controlled byte stream. 
Facilitated attacks, such as arbitrary code execution, have an impact on the confidentiality, integrity, or availability of the system. 
For instance, a \texttt{readObject} method present in a Java program is a weakness that is considered to be a vulnerability when: 
(1) the program containing this method accepts and deserializes data from a source that an attacker can control, and 
(2) the attacker can exploit this weakness. 
In practice (2) requires one to build a gadget chain and thus requires all necessary gadgets to be in classes that are on the classpath of the vulnerable application.
Note that classes containing gadgets do \emph{not need to be used} by the vulnerable program, just must be loadable.

\paragraph{\textbf{Gadget library.}}
By a gadget library, we denote a Java library containing one or more gadgets.
A gadget can be used during a deserialization attack when the corresponding gadget library is included in the classpath of the vulnerable application.  

\paragraph{\textbf{Exploit.}}
 An exploit is a piece of software or a sequence of commands that takes advantage of a bug or vulnerability causing a negative impact on the confidentiality, integrity, or availability of an impacted component or components.

 \paragraph{\textbf{Patch.}}
The  National Institute of Standards and Technology (NIST)~\cite{nistPatch} defines a patch as 
\emph{``a "repair job" for a piece of programming; also known as a "fix". A patch is the immediate solution to an identified problem that is provided to users''}. The patch of a library  \emph{``is not necessarily the best solution for the problem, and the product developers often find a better solution to provide when they package the product for its next release.''}

In the context of Java deserialization vulnerabilities, there are two kinds of patches: patches for vulnerabilities and patches for gadgets.
Patching a gadget library requires one to remove gadgets, thereby disabling attacks relying on this gadget library.
While authors of libraries sometimes decide to make exploitation harder by patching gadget libraries involved in attacks, 
gadgets are ultimately not flaws on their own.
{Most importantly, the gadgets are not themselves the deserialization vulnerabilities, they are rather pieces of code that facilitate the successful exploitation of a deserialization vulnerability that itself frequently resides in application code, not library code.}
One can draw a parallel with memory-corruption vulnerabilities.
Shacham~\cite{Shacham07} mentioned that libraries leveraged to exploit a memory corruption vulnerability are the ``innocent flesh on the bone''.
Thus, it is not surprising that they are rarely patched to address the issue. 
Patches for gadget libraries are similar to the heap-hardening~\citep{HallerKGB16, KuznetsovSPCSS14, Younan15} introduced in memory allocators in response to heap-based buffer overflow exploit techniques: 
they do not patch heap-based buffer overflows, yet they hinder exploitation techniques that have become publicly known.

\paragraph{\textbf{Deserialization attacks.}}
They are performed using two main steps: 
(1) an ahead-of-time \textit{serialization} step during which the attacker builds a customized byte stream specially crafted to execute a chain of gadgets during deserialization and 
(2) an online \textit{deserialization} step executed on the victim's vulnerable machine, and during which this victim's machine will deserialize objects from the attacker-controlled byte stream and thus execute the gadget chain. \\

\noindent \textit{Illustration.}  
Let us  use the example of Figure~\ref{example-terminology} to clarify the terminology. 
Assume that there is a library called \texttt{libA.jar} containing class \texttt{A} of Figure~\ref{fig:A-terminology}. 
As explained previously in the example of  section~\ref{sec:introduction}, the \texttt{command} field of type \texttt{String} of  class  \texttt{A} can be attacker-controlled. 
This class contains a \textit{gadget}: \texttt{Runtime.exec} (line 19)
 called through another gadget \texttt{readObject} of the same class \texttt{A}. 
The chain \texttt{A.readObject $\rightarrow$ Runtime.exec} is a \textit{gadget chain}. 
The \texttt{libA.jar} library contains these two gadgets and is then considered as a \textit{gadget library}.  
Now suppose that there is a program \texttt{VictimClass} using \texttt{libA.jar} in its classpath as shown in Figure~\ref{fig:B-terminology}. 
The \textit{vulnerability} in this context is that the code uses the \texttt{readObject} to deserialize objects from an untrusted source and that the gadget library \texttt{libA.jar} is in the classpath of \texttt{VictimClass}. An \textit{exploit} may consist in generating a file \texttt{f} by an attacker  in which he/she controls the field \texttt{command} and through which (i.e., when it is given as \texttt{args[0]} in line 6 of  \texttt{VictimClass.main()}) he/she can execute a command "\texttt{calc.exe}" for instance.  
The most secure \textit{patch} for this vulnerability is to not deserialize the untrusted file \texttt{f}. But, another possible patch consisting in  removing the class \texttt{A} from the classpath of the program \texttt{VictimClass} is sufficient to prevent this particular deserialization attack.

\begin{figure}[htbp]
\centering
\begin{minipage}{.46\textwidth}
\begin{myminted}[linenos=true,xleftmargin=5pt,fontsize=\scriptsize]{Java} 
/* Class A is in libA.jar */
public class A implements Serializable {
// the attacker can change the value of
// the 'command' field during serialization
  String command; 

// in class 'A', the 'command' field
// could be restricted to, e.g., a few values.
// This does not prevent the attacker from
// changing the field to a value of the attacker
// choice during serialization
[...]

  private void readObject() {
// this code is executed during deserialization
// if the attacker controls 'command', he/she 
// can execute arbitrary code on the machine
// deserializing an instance of type 'A'
    Runtime.exec(command); 
  }
}
\end{myminted}
\subcaption{A vulnerable class. (This sub-figure is a modified version of  Figure~\ref{fig:simple-example}.) }
\label{fig:A-terminology}
\end{minipage}
%\hfill
\begin{minipage}{.5\textwidth}
\begin{myminted}[linenos=true,xleftmargin=15pt, fontsize=\scriptsize, highlightlines={12}]{Java}
/* libA.jar is in the classpath of the 
   JVM running VictimClass.main() */
public class VictimClass implements Serializable {
  public static void main (String[] args) 
    FileInputStream fis = 
    	new FileInputStream(args[0]); 
    ObjectInputStream ois = 
    	new ObjectInputStream (fis);
// if the attacker gives an object of type 'A'
// to deserialize, A.readObject is executed
// before the cast to (String)
    String s = (String) ois.readObject();
    ois.close(); 	
  } 
}
\end{myminted}
\subcaption{A class deserializing an input file.} 
\label{fig:B-terminology}
\end{minipage}
\caption{Explanatory example for the terminology.}
\label{example-terminology}
\end{figure}

\subsection{Overview of Typical Deserialization Attacks} % or Insecure deserialization
\label{sec:deser-attack}

The requirement for this attack is that the victim machine runs software that deserializes objects from an untrusted byte stream controlled by the attacker.
In a first step, the attacker crafts a specific serialized file $s$ representing a class instance $i$.
Then the attacker sends $s$ to the victim either directly or through the network. 
Once received, the file is deserialized with a \texttt{readObject} method to try to reconstruct instance $i$.
The attack takes place during this deserialization process, when the Java code, 
relying on the attacker-controlled data in the byte stream, executes the attacker's payload. This payload then, for instance, may execute arbitrary code with the process' privileges through a call to \texttt{Runtime.exec()}.
An attack only works if the victim's Java process has all the required vulnerable classes on its \emph{classpath}.

Transforming a class instance $i$ into a byte stream is called serialization.
The basic principle of \emph{de}serialization is to rebuild the same class instance $i$ from the byte sequence. 
Figure~\ref{fig:B-terminology}  shows a short code snippet to illustrate the deserialization process in Java.
This generic Java code represents the software running on the victim machine. %in Figure~\ref{fig:attack}.
To simplify the code, the byte stream is read from a file and not from the network as in Figure~\ref{fig:B-terminology}.
The first and only argument passed to this program is a path to a file that  represents the serialized data to deserialize.
The Java code opens this file and calls the  method \texttt{readObject()} to deserialize its content (line 12 in Figure~\ref{fig:B-terminology}). 
Observe that while there is a cast to \texttt{String}, the attacker could put a different object type to deserialize.
Indeed, deserialization attacks are triggered during deserialization.
{The cast to the proper type (here \texttt{String}) is only executed \emph{after} the byte stream has been deserialized. 
Thus, the cast operation is executed too late and does not prevent an attack.}
 Observe also that there is no change of the methods \texttt{readObject()} and \texttt{writeObject()} signatures or code. 
 The only changes concern the fields accessible to the attacker.

\begin{figure}
\begin{center}
\begin{myminted}[linenos=false,xleftmargin=0pt,fontsize=\scriptsize, highlightlines={}]{Java} 
15 Runtime.exec()                           // (7)
14 NativeMethodAccessorImpl.invoke0()
13 NativeMethodAccessorImpl.invoke() 
12 DelegatingMethodAccessorImpl.invoke() 	
11 Method.invoke()                          // (6)
10 InvokerTransformer.transform()           // (5)
 9 ChainedTransformer.transform()           // (4) (d)
 8 LazyMap.get()                            // (3) (c)
 7 AnnotationInvocationHandler.invoke()     // (2) 
 6 $Proxy0.entrySet()                       // (2) (b)
 5 AnnotationInvocationHandler.readObject() // (1) (a)
 4 [...] // internal JVM code
 3 ObjectInputStream.readObject0() 
 2 ObjectInputStream.readObject()
 1 VictimClass.main(String[]) 
\end{myminted}
\end{center}
  \caption{Abstraction of call stack of the \texttt{Com\-monsCol\-lections1} attack.} 
 \label{CC1-Call-stack} 
\end{figure}

\subsection{A Concrete Real-World Example}

In this section, we describe the \texttt{Com\-monsCol\-lections1} deserialization attack (from  the ysoserial repository~\cite{ysoserial}).
More precisely, we analyze what happens  when the attack of \texttt{Com\-monsCol\-lections1} is exploited.
The gadgets are present in the 3.1, 3.2 and, 3.2.1 versions of apache commons-collections library. 

\begin{figure}[htbp]
\centerline{\includegraphics[scale=0.45]{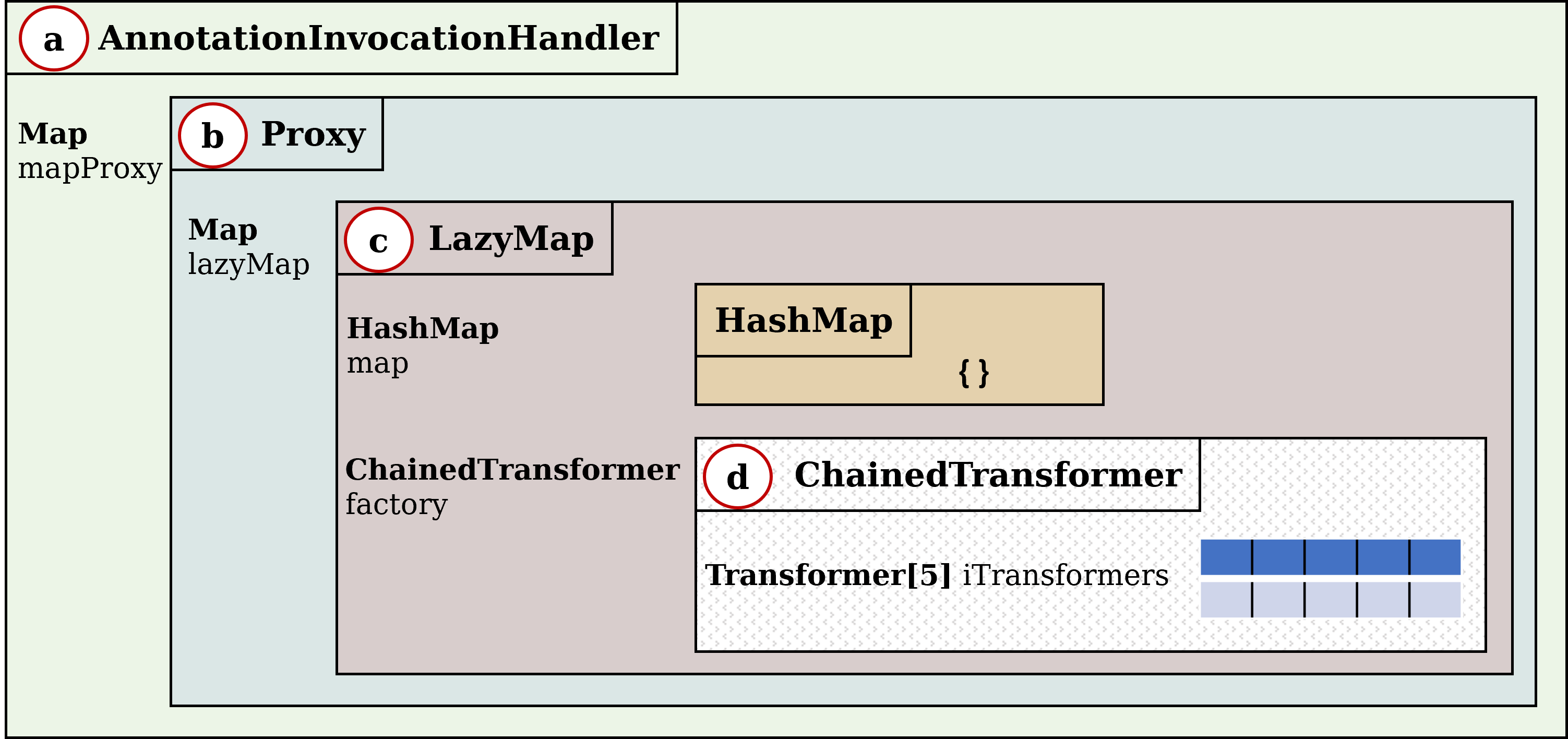}}
\caption{Components of the serialized byte stream generated by the ysoserial tool and which will be deserialized by the victim class in Figure~\ref{fig:B-terminology}.}
\label{fig:serializedFile}
\end{figure}

As shown in Figure~\ref{fig:serializedFile}, the serialized byte stream \texttt{s} generated for this attack features four main objects:
(a) an \texttt{An\-no\-ta\-tion\-In\-voca\-tion\-Handler}, 
(b) a \texttt{Proxy}, 
(c) a \texttt{LazyMap}, and 
(d) an array of \texttt{Transformers} (a \texttt{ChainedTransformer} object). 
The fields of each concrete object in the byte stream are controlled by the attacker.
During deserialization, the Java code might  choose to execute a branch based on the value 
of the fields controlled by the attacker. 
Furthermore, the Java code might call methods using fields controlled by the attacker.
The latter can thus control part of the execution flow.
In the \texttt{Com\-monsCol\-lections1} attack, objects (a-d) - of both Figures~\ref{CC1-Call-stack} and \ref{fig:serializedFile} - are \emph{chained} to trigger
the execution of the \texttt{Runtime.exec()} method with an attacker-controlled value
to achieve arbitrary code execution.
The call stack of this attack - when its payload is executed -  is represented in Figure~\ref{CC1-Call-stack}. 
%\is{
This call stack is composed  of a chain of \textit{gadgets}.  A gadget is a  method using objects or fields that can be attacker-controlled. 
To simplify Figure~\ref{CC1-Call-stack}, we do not consider the native calls related to Java  reflection  or the internal workings of the JVM. In the attack in this figure, we consider that there are 7 gadgets: (1) \texttt{AnnotationInvocationHandler.readObject()} which is the head, or the \textit{entry gadget}, of the chain, (2) \texttt{\$Proxy0.en\-try\-Set()}, (3) \texttt{LazyMap.get()}, (4) \texttt{ChainedTran\-sfor\-mer.trans\-form()}, (5) \texttt{In\-vo\-ker\-Trans\-for\-mer.transform()},  (6) \texttt{Method\-.in\-vo\-ke()} which we consider as an \textit{attack gadget} and (7) \texttt{Run\-ti\-me.ex\-ec()} representing the last gadget, the performed attack action.  An attack gadget is a method call  that triggers the payload. In this running example, \texttt{Method.invoke()} calls \texttt{Runtime.exec()} via reflection.

In this paragraph, we explain step-by-step the different calls in the gadgets chain. When the victim application, represented by the code in Figure~\ref{fig:B-terminology},
deserializes the \texttt{s} byte stream by calling \texttt{readObject} (stack frame 1 in Figure~\ref{CC1-Call-stack}), 
the internal JVM code handling the deserialization is executed (stack frames 2, 3 and 4). Range 4 represents 7 hidden method calls. We do not show them since they represent the internal JVM code related to Java reflection.
This leads to a call of the \texttt{readObject()} method of the first object to deserialize,
which is an \texttt{An\-no\-ta\-tion\-In\-voca\-tion\-Handler} (stack frame 5). 
In the code of this method, there is a field \texttt{this.memberValues} initialized by a \texttt{\$Proxy0}. This is the reason for calling \texttt{\$Proxy0.entrySet()} (frame 6). 

A Proxy in Java is a class generated during runtime to implement interfaces. It is associated with an invocation manager represented by the \texttt{In\-vo\-ca\-tion\-Han\-dler} class. 
The JVM uses reflection to redirect any method calling a Proxy to the  \texttt{in\-vo\-ke()}
method of the interface implemented by this Proxy. 
This explains the jump from \texttt{\$Pro\-xy0.en\-try\-Set()} to \texttt{An\-no\-ta\-tion\-In\-voca\-tion\-Hand\-ler.in\-voke()} (stack frame 7). This method \texttt{invoke()} will look for which method to call. This information is extracted from the serialized \texttt{s} byte stream in which there is a field initialized by the value \texttt{LazyMap}, a class of the commons-collections library. This value is assigned to a field called \texttt{this.memberValues} in the \texttt{invoke} method. This leads to the call of \texttt{LazyMap.get()} (stack frame 8). Until now, the \texttt{LazyMap} is empty. For this reason, the \texttt{get()} method uses a call to a \texttt{factory.transform()} method in order to decorate the \texttt{LazyMap}. Again here, the  \texttt{factory} field is extracted from the serialized \texttt{s} byte stream and is assigned to a \texttt{ChainedTransformer} leading to the call of  \texttt{ChainedTransformer.transform()} (stack frame 9). A \texttt{ChainedTransformer} implements the \texttt{Transformer} interface and contains four transformers: its first element is a \texttt{ConstantTransformer} which is equal to any constant chosen by the attacker, its three subsequent elements are \texttt{InvokerTr\-ans\-for\-mers} and each one of them will take the output of the previous one and transform it. The last \texttt{InvokerTransformer} will transform the attacker command (stack frame 10) into an invoke - by reflection -  of \texttt{Runtime.exec()} (stack frames 11 to 15).  This attack is using the Transformers objects. A \texttt{Transformer} is an interface in Java implemented by one or more other classes that transforms an input object into an output object. Among these classes are \texttt{ConstantTransformer} and
\texttt{InvokerTransformer}. Both of them are implementations of the \texttt{Transformer} interface.  The former allows to always return the same constant without checking the input object. The latter  allows the creation of a new object by reflection and the invocation of a method defined in the class of this object.

The commons-collections library has been patched in version 3.2.2. 
by adding a check on the \texttt{InvokerTrans\-former} object when calling its \texttt{transform()} method (line 10 of the call stack). 
A \texttt{check\-Un\-sa\-feSerialization} method checks whether serialization is enabled for unsafe classes like the \texttt{Transformer}. Otherwise, an exception is thrown. 
Observe that the library has been patched by changing the code of a single gadget (out of the four gadgets required for the attack). While this prevents \emph{this} attack, it does not prevent the reuse of the three untouched gadgets in future attacks. 
%-------------------------------------------------------------------------------

%-------------------------------------------------------------------------------
\section{Experimentation and Evaluation}
\label{sec:methodandeval}
\subsection{Methodology}
\label{subsec:methodology}

In this part, we explain how we proceed to perform the two types of analysis: on gadgets and on real-world Java applications. 

\subsubsection{Gadgets analysis}

\paragraph{Collecting the dataset.} 

Our experiments involve several elements: the ysoserial tool, libraries, and JVM versions. 

First, since we are referring to attacks that are using malicious files generated by the ysoserial tool, 
we download this tool~\cite{ysoserial}.
Second, we list all the libraries involved in the \NumberStudiedAttacks ~studied  ysoserial attacks. For  these attacks, there are \NumberOfStudiedLibraries ~ libraries involved. We download all the available versions of each library. In total, we have a set of \totalNumberOfLibrariesVersions ~jars for all \NumberOfStudiedLibraries ~libraries. Note that each library can have hundreds of versions. For example, there is an attack called \emph{Groovy1} which is using the Apache Groovy library. We have downloaded 192 versions of this library.  We have downloaded all the libraries' jars in September 2020. 
The third element of our study is the JVM.
They can be downloaded from the Oracle~\cite{Oracle},  IBM~\cite{IBM}, and the AdoptOpenJDK~\cite{AdoptOpenJdk}  websites. We obtained a dataset of \numberOfJVMs ~JVM versions containing 137 Oracle and 10 IBM  and OpenJdk versions. 

Our work aims at: 
(1) understanding how deserialization gadgets are introduced and patched in libraries and 
(2) collecting  the list of gadget library versions. 
The second goal is primordial in our work. Indeed, the studied ysoserial attacks are described for only some specific libraries and JVM versions, yet, we have discovered that most gadgets are still present in non-mentioned libraries/JVM versions.

\paragraph{Experimental Setup.}
\label{sec:setup}

In practice, we simulate a Java deserialization attack $A$ by following three steps:
(1) generate a malicious serialized file $MS$ corresponding to the specific attack $A$ using the ysoserial tool;  
(2) create an application including a victim class $V$ that deserializes the malicious input file $MS$ generated in step (1) using the  \texttt{Object\-Input\-Stream.read\-Object()} method; 
(3) add the gadget library(ies) i.e., libraries containing the required gadgets for the attack in the classpath and run the victim class $V$.

Note that even if the program deserializing data from an untrusted source does not directly \emph{use} classes containing gadgets, 
it is nonetheless vulnerable if these classes are on its classpath. This is because these classes can be referenced during deserialization. For instance, the code of Figure~\ref{fig:B-terminology} does not directly use the \texttt{LazyMap} class. However, we observe a call to the method \texttt{get()} on an instance of this class in the  call stack of Figure~\ref{CC1-Call-stack} (call frame number 8) since the commons-collections library (containing this class) is defined in the classpath of this victim code.\\
 
We combine the JVM version (an attack gadget might be present in a JVM)  with the libraries of the attack and run \numberOfExperiments~  programs. 
Consider an attack \textit{A} using a library \textit{l}. 
Suppose that \textit{l} has \textit{n} versions. For \textit{A}, we run \textit{$ \numberOfJVMs \times n$} executions. 
This means that we run each library version on all the \numberOfJVMs{} JVMs that we have collected. In another case,  an attack \textit{B} can use \textit{$nblibs > 1$} libraries. Suppose that each library \textit{l$_i$} of them has \textit{nbVers[l$_i$]} versions. The total number of executions is the sum of all the executions per library involved in the attack i.e., \[\numberOfJVMs \times nbVers[l_1] + \ldots +  \numberOfJVMs \times nbVers[l_i]+ \ldots +  \numberOfJVMs \times nbVers[l_{nblibs}]\]  
In other terms, for each attack  relying on $n$ library (n $>$ 1),  
we variate one library at once \textit{i.e.,} we fix ($n-1$) library versions that are vulnerable in the studied attack and we variate the versions of the remaining library.

 Table~\ref{tab:nb-experiments} explains where this ~\numberOfExperiments~ number  is coming from: it results from the addition of  all the numbers of executions per attack. For instance,  the CommonsBeanUtils1  attack relies on three libraries: commons-beanutils having 33 versions, commons-collections having 13 versions and commons-logging having 10 versions. For each one of these libraries, we did \textit{$ \numberOfJVMs \times n$} executions, where \texttt{n} defines the number of versions. At the end, we perform \textit{$ \numberOfJVMs \times 33 + \numberOfJVMs \times 13 + \numberOfJVMs \times 10 = 8232 $} executions for this attack. 
Our goal is to check if these attacks still are possible with these variants and understand what has changed to allow or block the attacks.

We developed a framework to automatically run our experiments and collect the results in log files correlated to each attack. 
For a single attack, there are thousands of log files. Our scripts consist of testing if the concerned combination (of the JVM and  the library(ies) versions) allows two actions to be performed: the serialization and the deserialization. 
If the serialization fails, the deserialization step cannot take place. 

We run all experiments on a machine with 12 x Intel(R) Xeon(R) Bronze 3104 CPU 1.70GHz, 256G of RAM, and the Debian 10.4 OS.

\begin{table*}[t]
  \centering
  \caption{Studied attacks and the number of experiments per attack}
  \resizebox{\textwidth}{!} {

\begin{tabular}{lllll|l}
\toprule
\textbf{Attack name} & \textbf{Lib name} & \textbf{\# lib versions} &  \textbf{\# experiments per lib} & \textbf{\# experiments per attack} & \textbf{Total}\\
\midrule
\emph{BeanShell1} & beanshell & 16 &  147 x 16 = 2352 & 2352 &  \multirow{27}{*}{\textbf{256515}} \\ \cline{2-5} \cline{1-1}
\emph{Clojure} &  clojure  & 145 &  147 x 145 = 21315 & 21315 &  \\
\cline{2-5} \cline{1-1}
\multirow{3}{*}{\emph{CommonsBeanUtils1}} & commons-beanutils & 33  & 147 x 33 = 4851 &  \multirow{3}{*}{4851 + 1911 + 1470 = 8232} & \\ 
\cline{2-4}
 &  commons-collections & 13 & 147 x 13 = 1911 & & \\
 \cline{2-4}
 & commons-logging & 10 & 147 x 10 = 1470 & & \\
\cline{2-5} \cline{1-1}
\emph{CommonsCollections1} &  commons-collections & 13 &  147 x 13 = 1911 & 1911 & \\
\cline{2-5} \cline{1-1}
\emph{CommonsCollections2} &  commons-collections & 13 &  147 x 13 = 1911 & 1911 & \\
\cline{2-5} \cline{1-1}
\emph{CommonsCollections3} &  commons-collections & 13 &  147 x 13 = 1911 & 1911 & \\
\cline{2-5} \cline{1-1}
\emph{CommonsCollections4} &  commons-collections & 13 &  147 x 13 = 1911 & 1911 & \\
\cline{2-5} \cline{1-1}
\emph{CommonsCollections5} &  commons-collections & 13 &  147 x 13 = 1911 & 1911 & \\
\cline{2-5} \cline{1-1}
\emph{CommonsCollections6} &  commons-collections & 13 &  147 x 13 = 1911 & 1911 & \\
\cline{2-5} \cline{1-1}
\emph{CommonsCollections7} &  commons-collections & 13 &  147 x 13 = 1911 & 1911 & \\ 
\cline{2-5} \cline{1-1}
\emph{Groovy1} & groovy & 192 &  147 x 192 =  28224 & 28224 & \\
\cline{2-5} \cline{1-1}
\emph{ROME} & rome & 12 & 147 x 12 = 1764 & 1764 & \\
\cline{2-5} \cline{1-1}
\emph{MozillaRhino1} &  js-rhino & 26 & 147 x 26 = 3822 & 3822 &  \\
\cline{2-5} \cline{1-1}
\emph{MozillaRhino2} &  js-rhino & 26 & 147 x 26 = 3822 & 3822 & \\
\cline{2-5} \cline{1-1}
\multirow{2}{*}{\emph{Spring1}} &  spring-beans & 180  & 147 x 180 = 26460 & \multirow{2}{*}{26460 + 27342 = 53802} & \\ \cline{2-4}
 & spring-core &  186  &  147 x 186 = 27342 & & \\
\cline{2-5} \cline{1-1}
\multirow{4}{*}{\emph{Spring2}} &  spring-core &  186  &  147 x 186 = 27342 & \multirow{4}{*}{27342 + 294 + 1470 + 27930 = 57036} &  \\ \cline{2-4}
 & aopalliance & 2 & 147 x 2 = 294 & & \\ \cline{2-4}
 & commons-logging & 10 & 147 x 10 = 1470 &  & \\ \cline{2-4}
 & spring-aop & 190 & 147 x 190 = 27930 &  & \\
\cline{2-5} \cline{1-1}
 \multirow{2}{*}{\emph{Click1}} & click-nodeps & 8 & 147 x 8 = 1176 & \multirow{2}{*}{1176 + 2940 = 4116}  & \\ \cline{2-4}
 & javax-servlet & 20 & 147 x 20 = 2940 &  & \\
\cline{2-5} \cline{1-1}
 \multirow{2}{*}{\emph{Vaadin1}} & vaadin-server & 199 & 147 x 199 = 29253 & \multirow{2}{*}{29253 + 29253 = 58506}  & \\ \cline{2-4}
 &  vaadin-shared & 199  & 147 x 199 = 29253 &  &\\
\cline{2-5} \cline{1-1}
\emph{JDK7U21} &  & 147 & 147 x 1 = 147 & 147 &   \\
\bottomrule
\end{tabular}
}
\label{tab:nb-experiments}
\end{table*}

\paragraph{Analyzing the results.}
\label{sec:res-analyze}
The results obtained from our experiments are  analyzed by:
\begin{itemize}
\item generating a table for each attack.  In each table, we have the versions of the JVM and of the implicated library(ies). Such table is composed of colored squares  with symbols: if the attack is successful the square is colored with red  and contains the \textit{0} number, otherwise, it is a fail. The failure of an attack might be caused by one of the three reasons: (1) the serialization is performed but the deserialization fails (orange color and the \textit{1} number)  or (2) the serialization fails because of "Unsupported major.minor version" (yellow color and \textit{V} symbol) or (3) the serialization fails because of an "Error while generating or serializing payload" generated by ysoserial (green color and \textit{-} symbol); 

\item then, filtering the results for the failed  attacks. Here distinguish between two reasons: 
either the serialization fails and there is no serialized file to read 
or the serialization succeeded but the deserialization fails.
\end{itemize}

The list of URLs used to download libraries used in the experiments as well as the tables generated by our experiments are all available at \url{https://github.com/software-engineering-and-security/java-deserialization-rce}.% \url{https://www.dropbox.com/sh/qu9hr6jv446zcja/AACtNVUCrlBl9ZGk-mdfG3dka?dl=0}. 

\subsubsection{Vulnerable-applications analysis}

Our second study consists in analyzing real-life Java applications containing deserialization vulnerabilities. 
Our goal is to study how vulnerabilities in these applications are patched.

\paragraph{Collecting the CVEs.}
To collect a suitable set of subject vulnerabilities, we searched specifically for Java deserialization vulnerabilities in the Mitre CVE database~\footnote{\url{https://cve.mitre.org/}}, using two queries: \emph{\{Java, deserialization\}, \{Java, deserialisation\}}.  
We found that there are \JavaDeserCVEs ~CVEs.

\paragraph{Analyzing the CVEs.}
Unfortunately, we observe that not all the CVEs resulting from our search are related to deserialization vulnerabilities.
Thus, we manually analyze the description of each CVE to classify them into one of the following three categories:

\begin{enumerate}
\item Deserialization Vulnerability (\emph{DV}): The kind of CVE we target in this paper, which describes an application in which there is a Java deserialization vulnerability (e.g., an attacker uses an entry point such as the \texttt{readObject} method in the application's code to deserialize his/her  untrusted data and carry out the attack).
\item GAdget (\emph{GA}): A CVE that describes a gadget, but not a vulnerability, i.e., there is no entry point for the attacker to carry out the attack.
\item Untrusted Code (\emph{UC}). A CVE that describes a vulnerability in the deserialization mechanism that can be exploited only if the attacker can execute arbitrary Java code.
\end{enumerate}

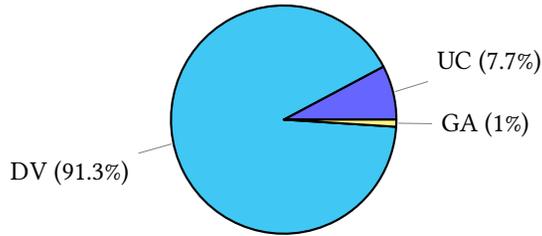
\begin{figure}%[htbp]
\centering

\begin{tikzpicture}

\pie[radius=1.5, text=pin, hide number]{7.7/UC (7.7\%), 91.3/DV (91.3\%), 1/GA (1\%)}

\end{tikzpicture}

\caption{More than 91\% of CVEs found via a search with keywords 
'Java' and 'deseriali[sz]ation' represent real Java deserialization vulnerabilities.}
\label{deser-categories} 
\end{figure}

%Table~\ref{tab:app-vuln} contains 104 CVEs which impact 65 applications. 
As represented in Figure~\ref{deser-categories}, we manually classified 
\DVCVEs ~(91.3\%) of these CVEs as DV, 
8   (7.7\%) as UC and 
1   ($\sim$ 1\%) as GA.
This means that the results of our search on Mitre with simple keywords contain noise (about 8\%) that we need to remove.

Table~\ref{tab:app-vuln} in Appendix~\ref{apps-vulnerable} shows a partial analysis of 29 CVEs for vulnerable Java applications.
The complete table with the \JavaDeserCVEs~ CVEs is available at \url{https://github.com/software-engineering-and-security/java-deserialization-rce}.%\url{https://www.dropbox.com/sh/qu9hr6jv446zcja/AACtNVUCrlBl9ZGk-mdfG3dka?dl=0}.  

\subsection{Experimental Evaluation}
\label{subsec:evaluation}

In this section, we address the following research questions: 
\begin{itemize}
\item RQ1: How Frequent are Deserialization Vulnerabilities? 
\item RQ2: How are Gadgets Introduced?
\item RQ3: How are Gadget Libraries Patched?
\item RQ4: What is the Life-cycle of Gadgets?
\item RQ5: How are Vulnerabilities Patched in Real-life Applications? 
\item RQ6: How easy is the automation of filters against deserialization attacks? 

\end{itemize}

\subsubsection{RQ1: How Frequent are Deserialization Vulnerabilities?}
\label{sec:rq_attack_frequency}

To understand the evolution of reported deserialization vulnerabilities, 
i.e., vulnerabilities in any programming language for which there is a CVE,
as well as the deserialization vulnerabilities specific to Java,
we conducted an empirical study based on the Mitre CVE database.

Deserialization vulnerabilities are widely spread in many languages. To understand the extent of such vulnerabilities in general, we firstly look for CVEs describing them.  A query with the "\emph{deserialization}" keyword in the Mitre's interface returns \MitreCVEs ~CVEs. 
We checked the alternative "\emph{deserialisation}" ($s$ instead of $z$) and found four matching CVEs all related to Java vulnerabilities. 
Out of these four CVEs, one is already present in the first search.
In total, we have identified \MitreCVEs{} + 4 - 1 =  \TotalCVEs  ~deserialization vulnerabilities.
The documented deserialization vulnerabilities were reported between 2004 and 2021. 
Among these, \numprint{15} are linked to the Apache commons-collections library (query \emph{deserialization, apache, commons, collections}).

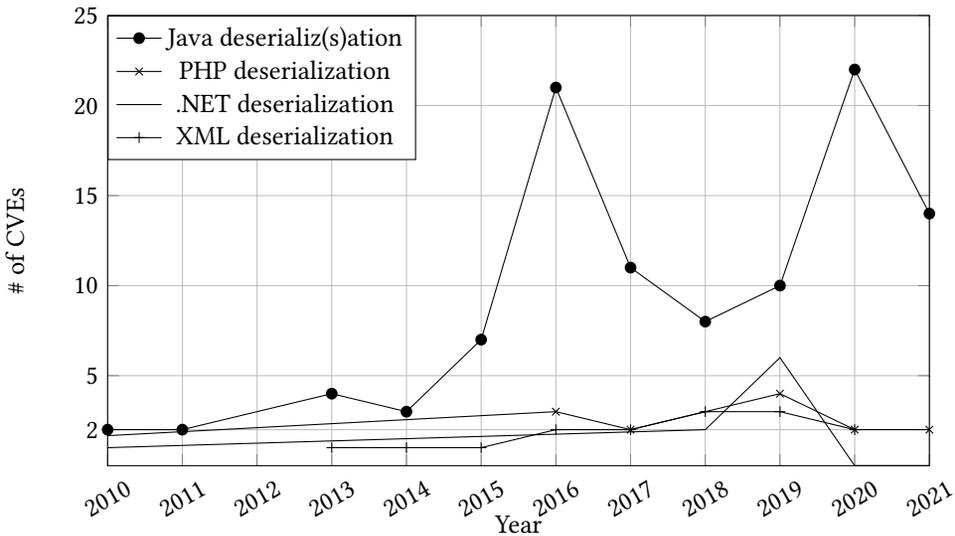
\begin{figure}[htbp]
\begin{tikzpicture}
	  \begin{axis}[	grid= major , xticklabel style={rotate=30, /pgf/number format/1000 sep=}, 
				width=0.9\textwidth,
				height=7.5cm,
				xlabel = {Year} ,
				ylabel = {\# of CVEs} ,
				xmin = 2010, xmax = 2021,
				ymin = 0, ymax = 25,
			 		xtick={2010,2011,2012,2013,2014,2015,2016,2017,2018,2019,2020, 2021},
				ytick={2,5,10,15,20,25},
				legend entries={Deserializ(s)ation, Java deserialization, PHP deserialization, .NET deserialization},
				legend style={at={(0,1)},anchor=north west}
				]
			
			\addplot[color=black,mark=*] coordinates {(2010,2) (2011,2) (2013,4) (2014,3) (2015,7) (2016, 21) (2017,11) (2018,8) (2019,10) (2020, 22) (2021, 14)}; 
			\addlegendentry{Java deserializ(s)ation}
			
			\addplot[color=black,mark=x] coordinates {(2004, 1) (2007, 1) (2016, 3) (2017,2) (2019,4) (2020, 2) (2021, 2)};
			\addlegendentry{PHP deserialization}
			
			\addplot[color=black] coordinates {(2010, 1) (2018,2) (2019,6) (2020,0) (2021, 0)};
			\addlegendentry{.NET deserialization}
			
			\addplot[color=black,mark=+] coordinates {(2013,1) (2014,1) (2015,1) (2016, 2) (2017,2) (2018,3) (2019,3) (2020,2)}; % Tracé point à point
			\addlegendentry{XML deserialization}
		\end{axis}
\end{tikzpicture}
\caption{Frequency of deserialization reported CVEs between 2010 and June 2021 according the MITRE database.}
\label{Deser-frequency-attacks} 
\end{figure}

Note that for searching CVEs in different languages, we need to use very specific keywords separated by space. However, checking  the results of our request is relevant to ensure that these results correspond exactly to what we look for. This is because the \textit{"results will include CVE Records that match all specified keywords"} as mentioned in the search tips of the Mitre website\footnote{\url{https://cve.mitre.org/find/search_tips.html}}. This means that the results of a request composed of two words \textit{"term1 term2"} may contain CVEs  concerning the vulnerabilities related to \textit{term1}, those which concern \textit{term2}, and those for both \textit{term1} and \textit{term2}. \\

The frequency of deserialization vulnerabilities encompassing Java programming language 
has a notable increase from 2015 until now. 
Indeed, we found that there are \JavaDeserCVEs ~CVEs among which \numprint{79} (76\%) were reported between 2015 and 2020, see Figure~\ref{Deser-frequency-attacks}.

We looked for further languages such as PHP, XML, and .NET and found that Java is the riskiest language for deserialization attacks among them. 
Java is one of the most popular and used languages 
(millions of developers run Java). This may explain why the percentage of (reported) vulnerabilities is higher than the other languages ~\cite{JavaSurvey}~\cite{GoJava}. 
Another point is the spike of vulnerabilities for Java in 2016. 
This might be related to Frohoff's research and his tool ysoserial.
The number of detected vulnerabilities by this tool helps other researchers to use them for detecting new attacks. 
For example, \emph{CVE-2016-2510} describes a gadget in the BeanShell  library for which the ysoserial tool presented an exploit called \emph{BeanShell1} at the beginning of 2016.

About 40 CVEs related to Java deserialization vulnerabilities have been reported between 2018 and 2020. 
 Most of Java deserialization vulnerabilities are critical because they allow arbitrary code execution on the victim machine.
This is probably one of the reasons Java programs and libraries are under scrutiny and so many vulnerabilities have been reported during the last five years.
Note that a single gadget may have several reported CVEs. 
This is caused by the fact that a gadget can be present in many applications and products. 
Then, a different CVE can be attributed  to each application for the same gadget.
Unfortunately, this kind of information is often not present in the description of CVEs which prevents us from
automatically counting unique deserialization vulnerabilities.

\myconclusion{
The overall trend shows that the number of CVEs related to deserialization is slightly increasing  in the last 10 years.
This means that in the real world, serialization is often used in applications to process untrusted data.
}

\subsubsection{RQ2: How are Gadgets Introduced?}
\label{sec:rq_how_vuln_introduced}

To answer  this research question, we analyze the results of the experimental protocol 
described in Section~\ref{sec:setup}. 
We consider that a gadget can be introduced in the library(ies) present in the classpath 
of a victim program and/or in the Java Class Library (JCL), the set of classes shipped with any JVM.

\paragraph{Introducing a gadget in an external library.}
Our objective is to analyze different attacks in order to show how libraries involved in these attacks include gadgets. Table~\ref{tab:detected-gadget-versions} shows our discoveries about gadget  library versions that were not mentioned in the ysoserial repository. The column  \textit{Discovered version} describes all the gadget library versions detected through our experiments. The last column (\textit{\# of new detected versions}) refers to the number of new gadget library versions not including the mentioned ones in ysoserial. For instance, it is mentioned, in this repository, that version 7.7.14 of vaadin-server library contains gadgets. Our experiments have found that  \textbf{135} more versions contain gadgets.

\begin{table*}[b]
  \centering
  \caption{Discovered versions of gadget libraries using our experiments}
  \resizebox{\textwidth}{!} {
%\footnotesize %\scriptsize
\begin{tabular}{p{0.25\textwidth}p{0.25\textwidth}p{0.5\textwidth}|p{0.25\textwidth}}
\toprule
\textbf{Library name} & \textbf{Version mentioned in ysoserial} & \textbf{Discovered versions} & \textbf{\# of new detected versions}\\
\midrule
beanshell & 2.0b5 & 2.0b4 and 2.0b5  & \textbf{1} \\
\hline
clojure & 1.8.0 &  1.6.0-beta1 until 1.9.0-alpha15 & \textbf{46} \\
\hline
commons-beanutils & 1.9.2 & 1.5 until 1.9.4 & \textbf{14} \\
\hline
commons-collections & 3.1 and 4.4.0 & 2.1.1, 3.0, 3.1, 3.2, 3.2.1, 3.2.2, 4.4.0-alpha1 and  4.4.0 & \textbf{6} \\
\hline
groovy & 2.3.9 & 2.3.0-beta-2 until 2.4.3 & \textbf{25}  \\
\hline
rome & 1.0 & 0.5 until 1.0 & \textbf{7} \\
\hline
js-rhino & 1.7R2 & 1.6R6, 1.6R7, 1.7R2 until 1.7.7 & \textbf{9} \\
\hline
spring-beans & 4.1.4.RELEASE & 3.0.0.RELEASE until 5.2.9.RELEASE & \textbf{140} \\ 
\hline
spring-core & 4.1.4.RELEASE & 4.0.0.RELEASE until 4.2.2.RELEASE & \textbf{21} \\
\hline
spring-aop & 4.1.4.RELEASE & 1.1-rc1, 1.1-rc2, 1.1, 3.0.0 until 4.2.9 & \textbf{64} \\
\hline
click-nodeps & 2.3.0 & 2.1.0-RC1-incubating until 2.3.0 & \textbf{6} \\ 
\hline
javax.servlet & 3.1.0 & 3.0.1 until 4.0.1 & \textbf{19} \\
\hline
vaadin-server & 7.7.14 & 7.0.0.beta1 until 7.7.17 & \textbf{135} \\ 
\hline
vaadin-shared & 7.7.14 & 7.4.0.beta1 until 8.11.3 & \textbf{122}\\
\bottomrule
\end{tabular}
}
\label{tab:detected-gadget-versions}
\end{table*}

For each library, we identify its first version containing gadgets and
the version just before, i.e., the version before introducing gadgets.
We then look at the log of the execution of the victim program using these two libraries, one per execution.
The log for the non-gadget library version contains a Java exception which explains the reason for the attack failure.
Using this technique, we have identified four surprisingly innocent-looking ways in which gadgets have been introduced:
(1) adding a class, 
(2) adding \texttt{java.io.Serializable} to the list of implemented interfaces, 
(3) adding a  method, and 
(4) making a class \texttt{public}.

Table~\ref{tab:intro-vuln} shows the studied attacks, the gadget library versions involved in these attacks, their versions without gadgets, and the actions performed to transform a non-gadget library into a gadget one.  Among the  19 studied ysoserial exploits, there are five  (26\%) that are relying on more than one library. These exploits are: \textit{CommonsBeanUtils1}, \textit{Spring1}, \textit{Spring2}, \textit{Click1} and \textit{Vaadin1}. 

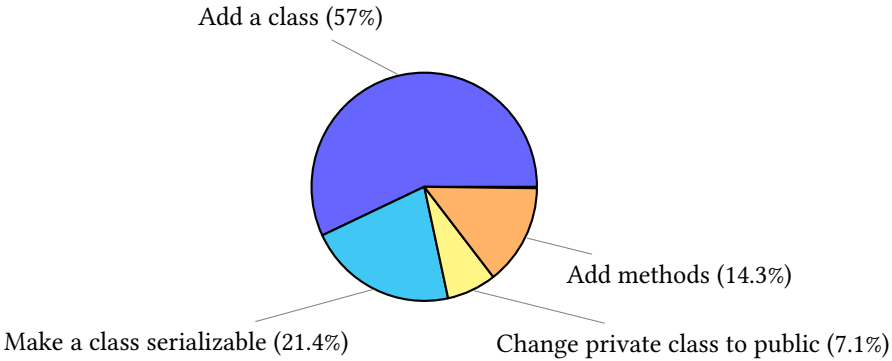
\begin{figure}
\centering
\begin{tikzpicture}
\pie[radius=1.5, text=pin, hide number]{57/Add a class (57\%), 21.4/Make a class serializable (21.4\%), 7.1/Change private class to public (7.1\%), 14.3/Add methods (14.3\%)}
\end{tikzpicture}
\caption{In more than 57\% of the \NumberOfStudiedLibraries ~studied libraries, gadgets are introduced by adding a class.} 
\label{pourcentage-intro-vuln} 
\end{figure}

As shown in Figure~\ref{pourcentage-intro-vuln}, there are eight versions (57\%) that integrate gadgets by adding a class (keyword  \emph{AddClass} in the figure), three (21.4\%) are affected by introducing the interface \texttt{java.io.Se\-riali\-za\-ble} in the list of their implemented interfaces (keyword \emph{MakeSerializable} in the figure). In the remaining cases, the first one introduces gadgets by changing the status of a class from \texttt{private} to \texttt{public} (7.1\%) (keyword \emph{ChangeToPublic} in the figure)
and two library versions introduce gadgets by adding new  methods  (14.3\%) (keyword \emph{AddMethods} in the figure). These methods are used for the construction of  malicious payloads.

\begin{table*}[t]
  \centering
  \caption{Actions performed to introduce a gadget in a library}
  \resizebox{\textwidth}{!} {
\begin{tabular}{p{0.25\textwidth}lp{0.25\textwidth}p{0.6\textwidth}}
\toprule
\textbf{Attack name} & \textbf{Gadget-free version} & \textbf{First gadgets version} & \textbf{Introducing gadgets action} \\
\midrule
\emph{BeanShell1} & beanshell-2.0b2 & beanshell-2.0b4 &  Change \texttt{private} class \texttt{bsh.XThis} to \texttt{public} \\
\hline
\emph{Clojure} &  clojure-1.6.0-alpha3  & clojure-1.6.0-beta1 &  Add  a class \texttt{AbstractTableModel\$ff19274a}  \\
\hline
\multirow{2}{*}{\emph{CommonsBeanUtils1}} & commons-beanutils-1.4.1 & commons-beanutils-1.5  & Add a class \texttt{BeanComparator} that implements \texttt{Serializable} \\
\cline{2-4}
 &  commons-collections (2001) & commons-collections2.1.1 & Add a class \texttt{ComparableComparator} \\
\hline
\emph{CommonsCollections1, 3 and 7} &  commons-collections3.0 & commons-collections3.1  &  Add \texttt{implements Serializable} to the class \texttt{LazyMap} \\
 \hline
 \emph{CommonsCollections2 and 4} &  commons-collections3.2.2 & commons-collections4-4.0-alpha1 & Add a class \texttt{TransformingComparator} to the library \\
 \hline
 \emph{CommonsCollections5 and 6} &  commons-collections3.0 & commons-collections3.1  &  Add \texttt{implements Serializable} to the class \texttt{TiedMapEntry} \\
 \hline
\emph{Groovy1} & groovy-2.3.0-beta1 & groovy-2.3.0-beta2 &  Add a  class \texttt{Opcodes}  \\
 \hline
\emph{ROME} & rome-0.4 & rome-0.5 & Add a  class \texttt{ObjectBean} \\
 \hline
 \emph{MozillaRhino1 and 2} &  js-rhino-1.6R5 & js-rhino-1.6R6  & Add a \texttt{private} method \texttt{accessSlot()} in the class \texttt{ScriptableObject}  \\
 \hline
\multirow{3}{*}{\emph{Spring1} and \emph{2}} &  spring-beans-2.5.6.SEC01 & spring-beans-3.0.0.RELEASE & Add a class \texttt{ObjectFactoryDelegatingInvocation\-Handle} which implements \texttt{Serializable}  \\ \cline{2-4}
  & spring-core-3.2.5.RELEASE &  spring-core-4.0.0.RELEASE  &  Add a class \texttt{SerializableTypeWrapper\$Method\-In\-voke\-TypeProvider} \\
 \cline{2-4}
  & spring-aop-1.0-rc1 & spring-aop-1.1-rc1 & Add \texttt{implements Serializable} to the class  \texttt{JdkDynamicAopProxy}\\
 \hline
 \multirow{2}{*}{\emph{Click1}} & click-nodeps2.0.1-incubating & click-nodeps2.1.0 & Add \texttt{implements Serializable} to the class \texttt{ColumnComparator}  \\ \cline{2-4}
 & javax-servlet & vulnerable from its first release & - \\
 \hline
 \multirow{2}{*}{\emph{Vaadin1}} & vaadin-server & vulnerable from its first release &  - \\ \cline{2-4}
 &  vaadin-shared-7.4.0-alpha14 & vaadin-shared-7.4.0-beta1 & Add a method \texttt{Capitalize(String)} in the class \texttt{SharedUtil}\\
 \hline
  \emph{JDK7U21} &  jdk1.7.0.25 & jdk1.6.0.04  & --  \\
\bottomrule
\end{tabular}
}
\label{tab:intro-vuln}
\end{table*}
  
In the following, we present two concrete exploits, \emph{BeanShell1} and \emph{CommonsCollections1}, and explain how gadgets have been introduced.\\

\noindent\textit{\bf\ul{BeanShell1.}}

The two gadget library versions are beanshell-2.0b4 and beanshell-2.0b5.
The version just before 2.0b4 is 2.0b2 and does not contain gadgets.
When looking at the content of the log file associated with the execution of the victim program using version 2.0b2, we observe that there is an error when serializing the payload \emph{BeanShell1}. 
This error is caused by an illegal access to a class \texttt{bsh.XThis}. 
We look for this class in the jar file  beanshell-2.0b2.jar. 
We found that this class is defined as \texttt{private}. 
Now, looking at this class in the gadget version beanshell-2.0b4, we found that  its definition has changed to \texttt{public class bsh.XThis}. This makes possible the access to this class, allowing  the serialization and thereby the attack.

\noindent\textit{\bf \ul{CommonsCollections1.}}
There are three gadget  versions of the library commons-collections: 3.1, 3.2, and 3.2.1 for this attack.
For  version 3.0, which does not contain gadgets, there is an error while serializing.
After analyzing the difference between the versions commons-collections3.0, which does not contain gadgets, and commons-collections3.1, which does contain gadgets, we found that the problem is originating from the class \texttt{LazyMap}. In  version 3.0, this class implements only \texttt{java.util.Map} and does not allow the serialization to be performed. In  version 3.1, it implements \texttt{java.util.Map} and \texttt{ja\-va.io.Se\-ri\-al\-iz\-able} allowing the serialization to succeed.

\paragraph{Detecting a flaw in the JCL.}

Deserialization vulnerabilities are not only found in Java code from third-party Java libraries but also in Java code from the JCL (Java Class Library) shipped with every JVM. Since this code is \emph{always} present in all JVMs, we discuss it in more detail here.
In our experiments, and for a studied attack, we vary the JVM versions in order to check if the attack succeeds or not for a gadget library version. Our goal is to identify the flaw in the Java Runtime Library (JRL) that allows the execution of such an attack.  We then filter the results and distinguish  three types of flaws allowing the insecure deserialization: (1) adding a \texttt{readObject()} method to a class of the JCL. This class can be used in the code to generate a malicious byte stream (example: the \texttt{BadAttributeValueExpException} class); (2) change the type of some fields to make them accessible and easily controlled by the attacker; (3) change the bytecode within a catch block.\\

\textit{Illustration}

\begin{figure}[tbp]
\begin{minipage}[t]{.45\textwidth}
   \centering
   \begin{myminted}[linenos=true,xleftmargin=15pt,fontsize=\scriptsize, highlightlines={10}]{Java}
private void readObject(...) throws 
  IOException, ClassNotFoundException {
     objectInputStream.defaultReadObject();
     AnnotationType instance;
     try {
         instance = AnnotationType.
               getInstance(this.type);
       }
     catch (IllegalArgumentException ex) {
        return;
       }
    [...]                
 }
 \end{myminted}
 \subcaption{readObject() in jdk1.7.0.21.} 
 \label{vuln-readObject}   
  \end{minipage}  
\hfill
  \begin{minipage}[t]{0.47\linewidth}
   \centering  
    \begin{myminted}[linenos=true,xleftmargin=15pt, fontsize=\scriptsize, highlightlines={10,11,12}]{Java}
private void readObject(...) throws 
  IOException, ClassNotFoundException {
    objectInputStream.defaultReadObject();
    AnnotationType instance;
    try {
        instance = AnnotationType. 
        	  getInstance(this.type);
        }
    catch (IllegalArgumentException ex) {
        throw new InvalidObjectException
        ("Non-annotation type in annotation
         serial stream");      
       }
    [...]                
 }
 \end{myminted}
\subcaption{readObject() in jdk1.7.0.25.}  
\label{not-vuln-readObject}  
  \end{minipage}
  \caption{Two versions of AnnotationInvocationHandler.readObject().}
  \label{fig:readObject-versions} 
\end{figure}

Consider the method \texttt{readObject()} of the class \texttt{An\-no\-ta\-tion\-In\-vo\-ca\-tion\-Hand\-ler} in the two JDK-versions jdk1.7.0.21 and jdk1.7.0.25, shown in Figure~\ref{fig:readObject-versions}.
When analyzing the two code snippets, it is important to note that the field \texttt{this.type} (used in line~7) might be attacker-controlled. The first \texttt{readObject} in jdk1.7.0.21 allows the deserialization attack because the catch block fails silently: it uses the instruction \texttt{return;} (line 10 of Figure~\ref{vuln-readObject}) which  will exit the \texttt{readObject} and let the attack continue. The \texttt{readObject} in jdk1.7.0.25 prevents the attack because the catch throws an \texttt{In\-va\-lid\-Ob\-ject\-Ex\-cep\-tion}  displaying a message \textit{"Non-annotation type in annotation serial stream"}. 
Even though the second version of \texttt{readObject} does not allow the attack, 
an analysis of the \texttt{try} block content i.e., the calls inside the \texttt{getInstance()} method, 
would be necessary to check for the presence of method calls that may invoke gadgets.

\myconclusion{
Java deserialization gadgets are not only introduced by adding new classes (8 out of \NumberOfStudiedLibraries ~libraries (57\%) added classes).
As code evolves, methods are updated or added, class signatures can change  (43\% of the \NumberOfStudiedLibraries ~libraries comprise such changes).
Thus, small code changes which look innocuous can frequently introduce gadgets.
}

\subsubsection{RQ3: How are Gadget Libraries Patched?} 
\label{sec:rq_how_vuln_patched}

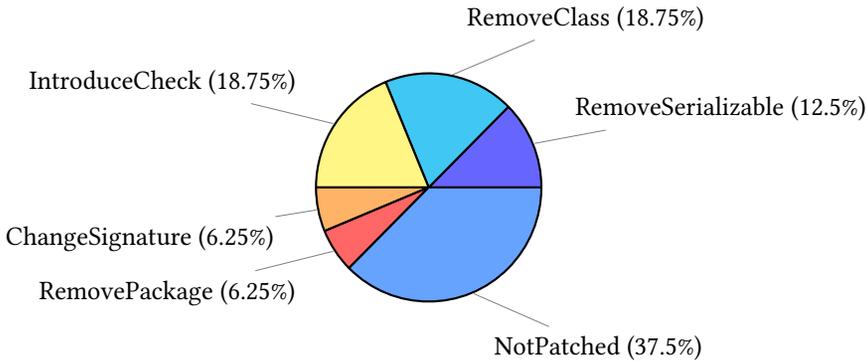
\begin{figure}[htbp]
\centering

\begin{tikzpicture}
%\footnotesize{
\pie[radius=1.5, text=pin, hide number]{12.5/RemoveSerializable (12.5\%), 18.75/RemoveClass (18.75\%), 18.75/IntroduceCheck (18.75\%), 6.25/ChangeSignature (6.25\%), 6.25/RemovePackage (6.25\%), 37.5/NotPatched (37.5\%)}
%}
\end{tikzpicture}
%}
\caption{Actions performed to remove gadgets from libraries.}
\label{pourcentage-patch-vuln} 
\end{figure}

Table~\ref{tab:patch-vuln} shows the actions performed to  fix flaws in the  studied gadget libraries. 
Note that the different patches present in this table can be either with the purpose of mitigating a gadget or a coincidental fix due to some changes for other purposes. In order to classify commits mitigating gadgets, we manually analyzed their messages. The result is represented in the last column of this table: when the action aims at fixing the gadget against deserialization, we mention it by "Y"; when it is a coincidental fix we put "N"; when the library is not patched, we use "-"; and when we do not find information about if the patch is intentional or coincidental we put the "Unknown" keyword.  
In this table, the second column represents the number of gadgets that we extract from each attack. 
 This number ranges from a minimum of 7 gadgets for the \emph{CommonsCollections1} attack to a maximum of 19 gadgets for the \emph{BeanShell1} attack. To find the corresponding patch in a library, we study the last version containing gadgets  and the version just after, which is gadget-free.
In this table, and as presented in Figure~\ref{pourcentage-patch-vuln}, a gadget patch may consist in: 
\begin{description}
	\item[{RemoveSerializable}]  removing the \texttt{java.io.Serializable} from the list of interfaces implemented by the vulnerable class. This action represents 12.5\% of cases,
	\item[{RemoveClass}] removing the vulnerable class in  18.75\% of cases,
	\item[{IntroduceCheck}] introducing a safety check  to  disable insecure serialization. This safety check can be an instruction in the code of a method or a whole added check method in 18.75\% of cases,
	\item[{ChangeSignature}] changing the signature of a method in 6.25\% of cases, or
	\item[{RemovePackage}] removing a package  from the gadget library (case of clojure) representing 6.25\% of cases. 
\end{description}

We have noticed that, in the \NumberOfStudiedLibraries ~studied libraries, only 8 of them are patched. We have identified 11 patching actions in this table. The remaining 6 libraries over \NumberOfStudiedLibraries ~(37.5\%) are not patched (\emph{NotPatched} keyword in the figure).

\begin{landscape}
\begin{footnotesize}
\begin{longtable}{p{0.2\textwidth}|p{0.03\textwidth}|p{0.15\textwidth}|p{0.03\textwidth}|p{0.17\textwidth}|p{0.32\textwidth}|p{0.2\textwidth}}
\toprule
\textbf{Attack} & \textbf{\rotatebox[origin=c]{90}{\shortstack[l]{Total \# \\ gadgets}}} & \textbf{Last gadget version} &\textbf{\rotatebox[origin=c]{90}{\shortstack[l]{\# gadgets \\ in lib}}} & \textbf{Gadget-free version} & \textbf{Applied patch}  & \textbf{\rotatebox[origin=c]{90}{\shortstack[l]{Intentional \\ patch (Y/N)}}}\\
\midrule
\emph{BeanShell1} & 19 & beanshell-2.0b5 & 15 & beanshell-2.0b6  & Remove the \texttt{java.io.Seria\-lizable} from the list of interfaces implemented by the class \texttt{XThis\$Handler} & Y (see link \href{https://github.com/beanshell/beanshell/commit/1ccc66bb693d4e46a34a904db8eeff07808d2ced\#diff-c3ddac5f59b0552b0cb83fca42e0f972dd20b01fefc0ecb6a999042cb43c44e0}{beanshellcommit}) \\
\hline
\emph{Clojure} & 10 & clojure1.9.0-alpha15  & 5 & clojure1.9.0-alpha16  & Remove the package "spec" used to load \texttt{clojure/spec/al\-pha\_\_init.class} or {clojure/spec/alpha.clj} from classpath  & N (see part 1.9.0-alpha16 of the link \href{https://clojure.org/releases/devchangelog}{clojurechangelog})\\
\hline
\multirow{2}{*}{\emph{CommonsBeanUtils1}} & \multirow{2}{*}{17} & commons-beanutils1.9.4  & 6 & not patched & not patched & - \\ 
\cline{3-7}
 &  & commons-collections3.2.2 & 0 & commons-collections4.4.0-alpha1  & Remove class \texttt{ComparableCom\-parator}  & N \\
\hline
\emph{CommonsCollections1} & 7 & commons-collections3.2.1 & 2 & commons-collections3.2.2 & Introduce additional constraint to  disable insecure serialization & Y (see link \href{https://commons.apache.org/proper/commons-collections/release_3_2_2.html}{collections3.2.2-release})\\
 \hline
 \emph{CommonsCollections2} & 13 & commons-collections4-4.0 & 2 & commons-collections4.4.1  & Remove the interface \texttt{java.io.Serial\-izable} from the list of implemented interfaces of the class \texttt{InvokerTrans\-former} & Y (see link \href{https://issues.apache.org/jira/browse/COLLECTIONS-580}{COLLECTIONS-580}) \\
 \hline
 \emph{CommonsCollections3} & 13 & commons-collections3.2.1 & 4 & commons-collections3.2.2 & Introduce additional constraint to  disable insecure serialization & Y (same as CommonsCollections1) \\
 \hline
 \emph{CommonsCollections4} & 15 & commons-collections4-4.0 & 4 & commons-collections4.4.1  & Remove the interface \texttt{java.io.Serial\-izable} from the list of implemented interfaces of the class \texttt{InstantiateTran\-sformer} & Y (see link \href{https://issues.apache.org/jira/browse/COLLECTIONS-580}{collections580}) \\
 \hline
 \emph{CommonsCollections5} & 8 & commons-collections3.2.1 & 5 & commons-collections3.2.2 & Introduce additional constraint to  disable insecure serialization & Y (same as CommonsCollections1) \\
 \hline
 \emph{CommonsCollections6} & 10 & commons-collections3.2.1 & 5 & commons-collections3.2.2 & Introduce additional constraint to  disable insecure serialization & Y (same as CommonsCollections1) \\
 \hline
 \emph{CommonsCollections7} & 9 & commons-collections3.2.1 & 4 & commons-collections3.2.2 & Introduce additional constraint to  disable insecure serialization & Y (same as CommonsCollections1) \\
 \hline
\emph{Groovy1} & 10 & groovy2.4.3 & 6 &  groovy2.4.4 & Add a check method called \texttt{readResolve()} in the class \texttt{MethodClosure} & Y (see  link \href{http://groovy-lang.org/security.html}{groovy}) \\
 \hline
\emph{ROME} & 15 & rome1.0 & 5 &  not patched & not patched & - \\
 \hline
\emph{MozillaRhino1} & 16 & js-rhino1.7.7.1  & 9 & js-rhino1.7.7.2 & Change the signature of the method \texttt{getSlot(String, int,int)}  to \texttt{getSlot(Object, int, int)} (consequence: not allowing the serialization of the malicious byte stream) & Unknown \\
\hline
\emph{MozillaRhino2} & 17 & js-rhino1.7.7.1  & 9 & js-rhino1.7.7.2   & Change the signature of the method \texttt{getSlot(String, int,int)}  to \texttt{getSlot(Object, int, int)} (consequence: not allowing the serialization of the malicious byte stream) & Unknown \\
\hline
\multirow{3}{*}{\emph{Spring1}} & \multirow{3}{*}{11} & spring-beans-3.0.0.RELEASE & 1 & not patched  &  not patched & - \\ \cline{3-7}
 & &  spring-core--4.2.2.RELEASE & 3 & spring-core-4.2.3.RELEASE & \texttt{readObject()} is instrumented with  an \texttt{Assert.state()} checking instruction & Y (see bug SPR-13656  in the link  \href{https://jira.spring.io/secure/ReleaseNote.jspa?projectId=10000&version=15296}{spring-io})\\ \cline{3-7}
 & & spring-aop-1.1-rc1  & 0 &  spring-aop-4.3.0.RELEASE  & Remove the class \texttt{DecoratingProxy} & Unknown \\
\hline
\multirow{3}{*}{\emph{Spring2}} & \multirow{3}{*}{12} & spring-core--4.2.2.RELEASE & 3 & spring-core-4.2.3.RELEASE & \texttt{readObject()} is instrumented with  an \texttt{Assert.state()} checking instruction & Y (see bug SPR-13656  in the link \href{https://jira.spring.io/secure/ReleaseNote.jspa?projectId=10000&version=15296}{spring-io}) \\ \cline{3-7}
 & & spring-aop-1.1-rc1  & 2 & spring-aop-4.3.0.RELEASE  & Remove the class \texttt{DecoratingProxy} & Unknown \\
\hline
\multirow{2}{*}{\emph{Click1}} & \multirow{2}{*}{17} & click-nodeps-2.3.0-RC1 & 5 & not patched & not patched & - \\ \cline{3-7}
&  & javax-servlet-api-4.0.1 & 0 & not patched & not patched & - \\
 \hline
 \multirow{2}{*}{\emph{Vaadin1}} & \multirow{2}{*}{10} & vaadin-server-7.7.17 & 2 & vaadin-server-8.0.0  & Remove the class \texttt{PropertysetItem} from the package
\texttt{com.vaadin.data.util} & N \\ \cline{3-7}
 & & vaadin-shared-7.4.0.beta1  & 0 & not patched &  not patched & -  \\
\hline
JDK7U21 & 11 & jdk-1.7.0.21 & 11 & jdk-1.7.0.25 & Add check (try/catch bloc) in the \texttt{readObject} method & Unknown \\
 \bottomrule
\caption{Different actions to fix a flaw in libraries}
  \label{tab:patch-vuln}
\end{longtable}
\end{footnotesize}
\end{landscape}

Let us now analyze two different actions performed to fix the  attacks \emph{BeanShell1} - described in Section~\ref{sec:rq_how_vuln_introduced} - and \emph{CommonsBeanUtils1}. The patch of the libraries involved in these two attacks is described in Table~\ref{tab:patch-vuln}.  \\

\noindent\textit{\bf \ul{BeanShell1.}}
This gadget library is patched in the version \emph{beanshell-2.0b6} by removing  \texttt{java.io.Se\-ria\-li\-za\-ble} from the list of interfaces implemented by the class \texttt{XThis\$Handler}. 
As a result, this class can no longer be serialized. 
\\

\noindent\textit{\bf \ul{CommonsBeanUtils.}}
First, note that this attack uses gadgets in  two libraries: commons-beanutils and commons-collections. 
We found that the gadget library commons-beanutils was never patched.
For the commons-collections library, the gadget is patched by removing the \texttt{Com\-parable\-Com\-parator} class.

\myconclusion{ 
Once a class is in a library, removing it might break backward compatibility, thus this removal is not often an option.
Among the studied patches, 18.75\% remove a class and 18.75\%  add a check.
Other solutions include adding a safety check to disable the insecure deserialization, removing the \texttt{java.io.Serializable} interface from the list of implemented interfaces or even removing a whole library package. 
A significant number of cases (37.5\%) are not patched at all.
}

\subsubsection{RQ4: What is the Life-cycle of Gadgets?}
\label{sec:rq_vuln_lifecycle}

To define the life-cycle of a gadget library or a gadget library version, we extract the following dates:
\begin{itemize}
\item when the first library version has appeared, i.e., the jar file appearance date,
\item when the gadget was introduced, and
\item when the library was patched. We look either at the appearance date of the released version free from gadgets or at the date of the patch in CVE if it exists.
\end{itemize} 

\begin{figure*}
\centerline{\includegraphics[width=\textwidth]{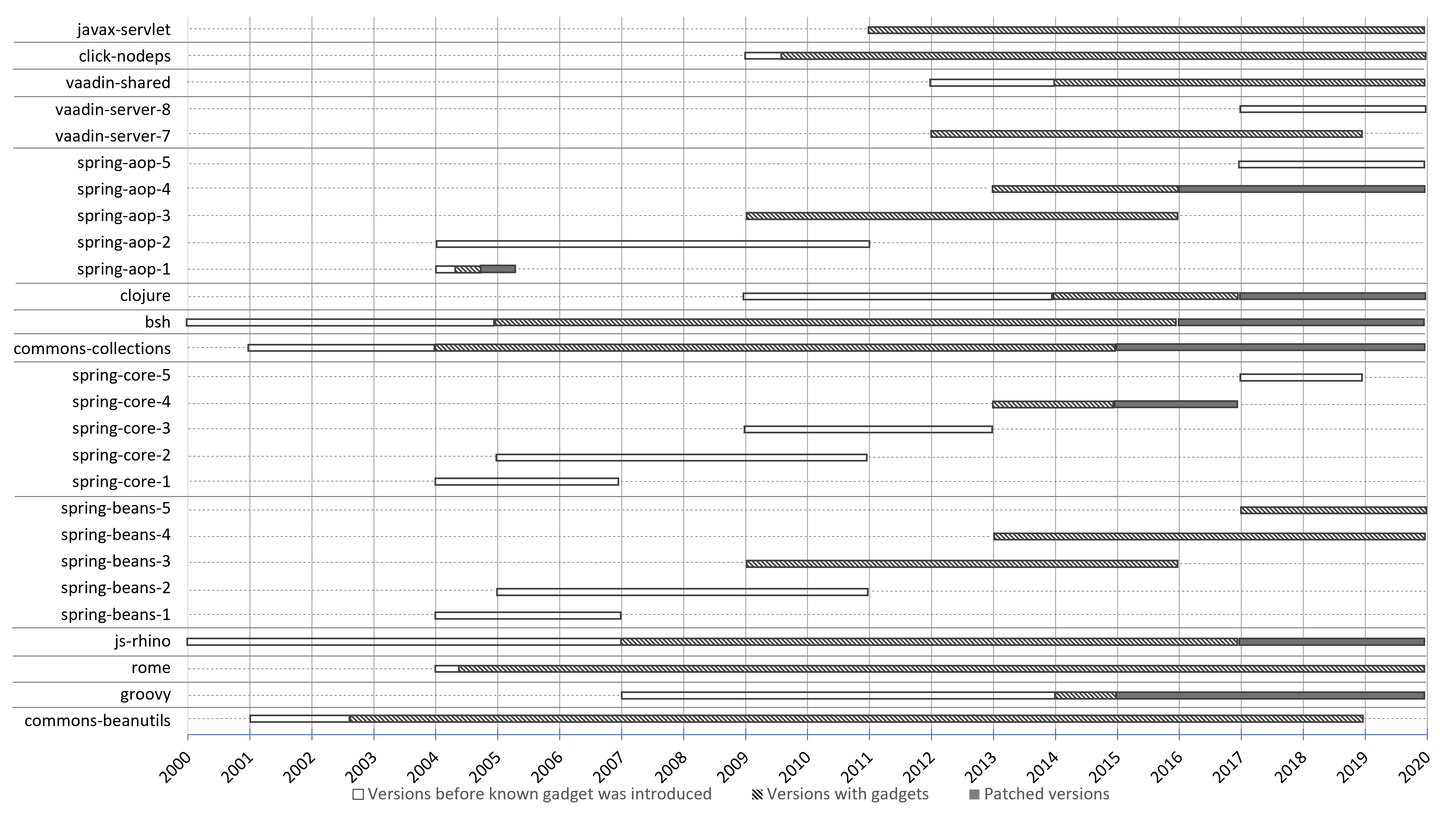}}
\caption{Life-cycle of some library versions.}
\label{patching-time} 
\end{figure*}

Figure~\ref{patching-time}  shows the answers to  this RQ according to the different studied attacks. 
  In this figure, each line represents the life-cycle of a library for which  we distinguish between: (1) the versions before the known gadgets were introduced (uncolored rectangle); (2) the  versions that  contain gadgets (dashed rectangle);  and (3) the patched versions of the gadget library (black rectangle). 
  First of all, note that the libraries  spring-beans, spring-core, spring-aop, and vaadin-server have several versions developed in parallel. Each one of these versions is described by a line in the figure. % For this reason, we separe them in different lines. 

According to Figure~\ref{patching-time},  we observe that most of the studied attacks are detected late. In fact, for some libraries like groovy and js-rhino, the detection of the gadgets takes 7 years.  This can be explained by the fact that the ysoserial tool has been developed in 2016 and the introduced vulnerabilities use versions of libraries that were developed from the beginning of the 2000s.  
We also observe that most of the libraries take between two months and several years to be patched. 
The longest time is for the commons-collections and bsh libraries and is equal to eleven years. 
The shortest time is two months for the  library spring-aop-1.   
Furthermore, we have classified the studied (versions of) libraries into three categories: 
\begin{enumerate}
\item[\textit{Cat1}] (versions of) libraries for which gadgets were introduced and then patched. This category contains the groovy, js-rhino, commons-collections, bsh, clojure, and spring-aop-1 libraries.
\item[\textit{Cat2}] libraries which are never patched, like the commons-beanutils, rome, spring-beans, vaadin-server-7, vaadin-shared, click-nodeps,  and javax-servlet libraries.
\item[\textit{Cat3}] library versions that contain gadgets from their appearance date.  We cite as examples the vaadin-server-7, spring-aop-3, spring-aop-4, spring-beans-3, spring-beans-4, and spring-beans-5.
  
\end{enumerate}

\noindent Observe that the presentation of the spring-core-4 library is simplified in the Figure~\ref{patching-time}. 
Indeed, this version contains four branches: 4.0.* \ldots 4.1.* \ldots 4.2.* and 4.3.* which are developed in parallel. 
The gadgets were initially patched in one branch and, a few months later, ported to other branches.
Merging all these branches in spring-core-4 would have resulted in a ``contains gadget'' ``patched'' ``contains gadget'' ``patched'' pattern which is wrong since it suggest the gadgets are reintroduced.
To simplify the figure, we represent spring-core-4 as being patched when the first branch is patched and ignore the fact that there is a delay to propagate this patch to other branches.  

Note that our study about the patching time of gadgets does not aim to  advocate fixing patches so quickly. 
Rather than that, our goal is to give an observation about the time that is taken to patch existing gadgets. 

\myconclusion{
When introducing  gadgets in a library, the latter can be patched in few  months. However, for some libraries, the patch is applied after several years going until more than 10 years. 
}

\subsubsection{RQ5: How are Vulnerabilities Patched in Real-life Applications?}
\label{sec:rq_apps}

For this study, we use the \DVCVEs ~CVEs 
describing Java deserialization vulnerabilities we have identified 
in Section~\ref{sec:rq_attack_frequency} (see Figure~\ref{deser-categories}).
We manually analyzed their corresponding \NBApps ~applications containing the vulnerabilities
to understand what has been changed in these applications to prevent the exploitation
of these vulnerabilities.

We observe that 69 applications (89.6\%) have a single CVE, yet that eight applications (10.4\%) have multiple reported CVEs. 
For example, the Atlassian Bamboo before 5.9.9 and 5.10.x before 5.10.0 application had two CVEs in two years:
First CVE-2014-9757, for which the fix consists in upgrading the Smack library used by the Bamboo application. 
Unfortunately, this fix does not prevent the next vulnerability, CVE-2015-8360.
The fix for this second CVE is a patch that introduces both allow and deny lists.
Thus, even though a fix exists, it may be not sufficient to protect the application from future attacks.

Among all the \DVCVEs ~CVEs associated with the \NBApps ~studied applications, we successfully analyzed \SuccessfulCVES ~CVEs based on:
the CVE description (41 ~CVEs), 
the code of the impacted application (6 CVEs),
or the workaround description (11 CVEs).
We were unable to analyze 37 CVEs either because the code of the applications is not publicly available (neither source nor bytecode, for 36 CVEs) or 
because our manual analysis exceeded a time limit (1 CVE).
A description of 29 CVEs among the \DVCVEs ~CVEs is available in Table~\ref{tab:app-vuln} of Appendix~\ref{apps-vulnerable}.

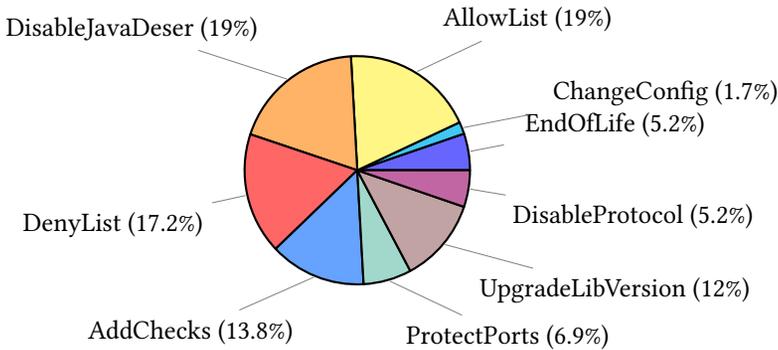
\begin{figure}[htbp]
\begin{minipage}{.95\textwidth}
\centering
\begin{tikzpicture}%\scriptsize
\pie[radius=1.5, text=pin, hide number]{
5.2/\phantom{a}EndOfLife (5.2\%), 
%3.7/\phantom{aaaa}UnknownInaccessible (3.7\%),
1.7/\phantom{aaaa}ChangeConfig (1.7\%),
19/AllowList (19\%),
19/DisableJavaDeser (19\%),
17.2/DenyList (17.2\%), 
13.8/AddChecks (13.8\%),    
6.9/ProtectPorts (6.9\%), 
12/UpgradeLibVersion (12\%),
5.2/DisableProtocol (5.2\%)
}
\end{tikzpicture}

\caption{Patching/Mitigation actions for the \SuccessfulCVES~  CVEs.} 
\label{patching-categories} 
\end{minipage}
\end{figure}

In total, we were  able to analyze \SuccessfulCVES ~CVEs from 52 applications.
Figure~\ref{patching-categories} presents the different categories of patches and mitigation techniques software vendors have devised.  
Among the \SuccessfulCVES~solutions, 
eleven (19\%) disable (de)serialization functionalities (e.g., CVE-2020-11973);
eleven (19\%) add an allow list containing the list of allowed classes or packages to use deserialization (e.g., CVE-2013-2165);
ten (17.2\%) add a deny list in which the classes are not allowed to be deserialized (example of CVE-2018-20732);
eight (13.8\%) add checks in serializable classes (keyword \emph{AddChecks} in the figure) 
which may go from adding some instructions (e.g., CVE-2016-6793) to activating the sandbox (e.g., CVE-2018-1000058);
seven (12\%) upgrade library versions (e.g., CVE-2014-9757);
four (6.9\%) protect or restrict access to ports  (e.g., CVE-2017-10934);
three (5.2\%) disable protocols (e.g., CVE-2015-4852);
three (5.2\%) do nothing because the application software reaches the end of life (e.g., CVE-2016-7065);
one (1.7\%) changes the software configuration (CVE-2020-9493).

Note that for  13  CVEs there is no patch (i.e., the code is unmodified) but only a workaround solution. 
Such methods consist mainly in blocking  access to the vulnerable code in order to reduce the severity of the impact of the attack or to prevent it. They do not modify the vulnerable code itself but work around this code to prohibit the access for performing the attack. 
For instance, the solution to mitigate CVE-2018-15381 consists in blocking or protecting the access to a port. 
Obviously, this does not remove the vulnerability in the application in question. 
In other words, the application is still exposed to the risk of attacks if, for some reason, the access to the port becomes allowed again
(e.g., new software configuration, software deployed in a new environment).

We complete our analysis by the study the nature of commits patching deserialization vulnerabilities:
are these patches manual code modifications or are they automatically generated by a tool?
To have access to commits/patches, the applications must be open-source. Thus, we selected 25 open-source vulnerable applications which use deserialization. 
Among these 25 applications, listed in Table~\ref{tab:patches-commits},
we found 17 applications having code patches that were manually generated and described in commits.
The remaining seven applications either update an external dependency (4), are end-of-life (1) or we were unable to find the patch (2). 
We observe that no commit is classified as being generated by a tool. 
The results of our analysis  are described in Table~\ref{tab:patches-commits} of Appendix~\ref{analyze-commits}.

\myconclusion{
Deserialization vulnerabilities may be present on several occasions for the same application. Disabling the deserialization in Java applications and building an allow-list are the most popular kinds of patches for this kind of attacks (19\% of cases for each one of these actions).  Such patches may prevent deserialization attacks once and for all. However, for some applications, the patching/mitigation action consists in upgrading library versions (12\% of cases), which is not safe since the attacker could find another way to carry out the attack.
All patches in the open-source applications we have analyzed were manually written which may suggest that tools are rarely used to automatically fix this kind of vulnerability.
}

\subsubsection{RQ6: How easy is the automation of filters against deserialization attacks?}
\label{sec:rq_filters}

Since  2016, Java supports a filtering mechanism to prevent the exploitation of deserialization vulnerabilities.
This filter, described in JEP 290~\citep{jep290,redhatJdk}, can restrict classes to be deserialized to a specific set (allow list) or 
prevent a specific set of classes from being deserialized (deny list).

To use these filters, we have to extract classes used by Java applications to either:
(1) create a deny list containing at least one class from each deserialization attack (in order to prevent each attack) 
and make sure that the deny list will not break any Java application or
(2) create an allow list containing all classes used by the Java application during deserialization
and make sure that it prevents all the attacks.
To be able to create these allow and deny lists, we first extract the set of classes required by the \NumberStudiedAttacks ~ysoserial deserialization attacks.
These classes are listed in Table~\ref{tab:type-extract}.
We then manually extract the set of classes used during deserialization from 10 real-world vulnerable applications which use deserialization. 
This set is listed in Table~\ref{tab:apps-type-extract}. 

Unfortunately, we observe that all the Java applications make use of generic types such as \texttt{ArrayList<Object>} or \texttt{Serializable}.
This is problematic since the allow list should then contain all possible serializable classes 
and the deny list should then be empty which makes the filters useless.
This shows that an analysis of the code deserializing data is not precise enough to generate a useful set of classes that can be used in filters.
There are two solutions that can be explored as future work.
The first is to rely on the in-depth knowledge of the applications by the developers themselves.
However, in practice, it might be difficult for the following reasons: 
contacting the developer responsible for the code might not be possible because the person might have left the project; 
the developer might not remember precisely how the code works, etc. 
The second is to automate the analysis of the applications to automatically extract the set of classes used by the application.
While at a first sight this approach seems to work, it will face the challenges of static analysis such as reflection, code loading, and other language-specific features difficult to analyse automatically.

Furthermore, as the code of applications evolves, the allow/deny lists must be kept synchronized with the new versions.
This would require that developers spend time debugging and updating these lists instead of working on the application's code directly.

\myconclusion{
In the lab, deserialization attacks can be trivially blocked when activating the filter.
However, in practice setting up and maintaining these filters suppose that  developers  know in advance the types to deserialize in the application which, often, is not trivial because of technical limitations, time constraints, or project management challenges. 
}

%-------------------------------------------------------------------------------

%-------------------------------------------------------------------------------
\section{Synthesis and Take-away Messages}
\label{sec:synthesis}

Our work has yielded the following important conclusions:  (1) the persistent deserialization of untrusted data; (2) what is the effect of disabling the \texttt{java.io.Serializable} in the patches; (3) what are the factors impacting the duration of finding a patch;  (4) how precise is the definition of deserialization vulnerabilities in CVEs description; and (5) what are the take-away messages from our analysis results.

\subsection{Persistent deserialization of untrusted data}
The deserialization of untrusted data is not recommended by the  OWASP: "the only safe architectural pattern is not to accept serialized objects from untrusted sources or to use serialization mediums that only permit primitive data types."~\cite{owaspDeserPrevention}~\cite{hdivDeserPrevention}. The same advice is provided by the Secure Coding Guidelines for Java SE which  states,  at the beginning of its 8th section, that "Deserialization of untrusted data is inherently dangerous and should be avoided"~\cite{SCGJSE2020}. However, our analysis of Java applications shows that, until now, developers are still using deserialization of data from untrusted sources.
This could be explained by developers not being aware of the recommendations, a lack of proper security vetting mechanisms during software development,
or constraints imposing the use of insecure legacy code.

\subsection{Disable Serializable}

When analyzing the gadget patches (see Table~\ref{tab:patch-vuln}), we find that disabling  the deserialization, by removing the \texttt{java.io.Serializable} interface from the list of implemented interfaces in the application's class(es), prevents exploitation in every single case. Removing this interface is very effective to protect against the known  deserialization attacks  since it breaks the chain of gadgets in the victim application at hand. The other solutions, such as changing the library versions in dependencies, or blocking the access to some ports, can significantly reduce the risk of exploitation but do not remove the weakness.

\subsection{Library use frequency vs. duration of finding patches}
A large fraction of libraries,like the click-nodeps and the javax-servlet, or applications is never patched. Some libraries, like  the commons-collections3, have been  deprecated and replaced by other releases (collections4 in the case of commons-collections). In the other cases, the patch takes many years to be carried out. This points to an important question: Is there a link between the frequency of using a library in real-world applications and the duration to fix its gadgets? In other words, if a library containing gadgets is used in many applications, do developers find the gadgets and fix them more quickly than for other unsafe libraries that are only infrequently used?  We have found a first element to answer this question: the frequency of using a library in applications is not the main factor impacting the duration of finding a patch. For example, commons-collections is used in many real-world applications, yet it was unsafe between 2004 and 2015 and only then, after 11 years, the patch was introduced to fix its gadgets in both deprecated version 3.2.2 and its release 4.4.1.

\subsection{Definition of deserialization vulnerabilities in CVEs}
The definition of deserialization vulnerabilities in CVEs is not precise. In fact, we have found that some declared deserialization vulnerabilities CVEs are in reality not vulnerabilities but rather  descriptions of gadgets (there is no entry point for the attacker to carry out the attack) or descriptions of untrusted code (vulnerability in the deserialization mechanism which can be exploited only if the attacker can execute arbitrary Java code).

\subsection{Gadgets inspection}
A usable take-away from our analysis results is the list of recommendations library developers should follow to prevent the introduction of gadgets. 
These recommendations are inferred from Table~\ref{tab:intro-vuln}. 
When a developer introduces a new class that can be serialized or modifies an existing one (see Table~{\ref{tab:intro-vuln}} for real-world examples), we suggest the following recommendations:
\begin{itemize}
 \item List all the new field types and sub-types which are introduced or modified.
 \item For each type, check that there is no code reachable from the \texttt{readObject} method which enables either to (1) jump to a known gadget or (2) execute code based on the untrusted data (e.g., a reflective method call with the method description extracted from the untrusted input).
\end{itemize}

%-------------------------------------------------------------------------------

%-------------------------------------------------------------------------------
\section{Limitations}
\label{sec:limitations}

\subsection{Scope of our study}

Our work is based on vulnerabilities described by the ysoserial tool. 
In fact, the generation of the serialized files was conducted by the commands described in its repository. This tool is developed and widely adopted by security researchers due to its ease of use for creating proof of concept payloads. It gathers a wide range of exploits that  concern many well-known libraries such as the commons-collections library. We are not aware of any other tools with features comparable to ysoserial.
Yet, the scope of our study is limited to the gadgets described in ysoserial, and as result, we cannot affirm the complete list of gadget libraries. 
Furthermore, the list of actions performed to introduce gadgets in the libraries in Table~\ref{tab:intro-vuln} is not exhaustive. These actions are extracted from the \NumberStudiedAttacks ~exploits described in ysoserial. 

\subsection{Source code accessibility}

When performing our analysis of the vulnerabilities in Java applications, we encountered some difficulties in finding and obtaining access to the source code of some applications. Once found, the code is sometimes hard to analyze, and finding the patch takes a lot of time. Another limitation regards the inaccessibility of some existing patches for 4 CVEs among our studied \DVCVEs~ ones (4.1\%). Indeed, some applications such as IBM Maximo Asset Management (CVE-2020-4521) present links to patches that are not accessible: an error message is displayed when trying to obtain access to the patch. It is not trivial to extract this patch information in these cases and a reverse engineering step is needed to get such information. For time-related reasons, we could not do this step. Obviously, this harms our study since we cannot identify supplementary patching actions that may be unknown beforehand.   

\subsection{CVEs keyword-based search precision}
The study on the frequency of deserialization vulnerabilities yields an under-approximation of the real-world situation.
Indeed, some CVE might not have the keyword \emph{deseriali[sz]ation} we use for the search and thus might not be present in
the list of CVEs returned from the search.
This is the case for instance for CVE-2021-26858~\footnote{\url{https://cve.mitre.org/cgi-bin/cvename.cgi?name=CVE-2021-26857}} a .NET deserialization vulnerability used by the HAFNIUM group to run code as SYSTEM on Microsoft Exchange servers~\cite{microhafnium}: the description of this vulnerability on Mitre's website does not mention that it is a deserialization vulnerability.
Such CVE descriptions were incomplete at the time of writing but might be updated to be more precise.

%-------------------------------------------------------------------------------

%-------------------------------------------------------------------------------
\section{Related work}
\label{sec:state-art}

\subsection{Deserialization vulnerabilities}

Preventing deserialization attacks starts right at the serialization step.
As explained in~\cite{LawrenceBestPractices}, at this stage it is important to follow recommendations and best practices for the secure use and implementation of Java serialization. Some tips are given to make the code more secure such as (1) guard sensitive data fields; 
 (2) check all security permissions for serialization and deserialization carefully and (3) use serialization filtering for untrusted data. 
The detection of this kind of vulnerability may be attempted early in the software development process.
In this context, Koutroumpouchos \textit{et al.}~\cite{Koutroumpouchos19} have  developed a dynamic tool called \textit{ObjectMap} allowing the detection of deserialization  and object injection vulnerabilities in Java and PHP-based web servers. 
Their tool accepts as inputs a URL of such a web server and generates HTTP requests containing payloads of known attacks. 
Executing these requests enables to detect if the web server is vulnerable to known attacks. 
If one of the payloads is executed it means that a vulnerability has been executed and that the web server needs to be patched.

Frohoff \textit{et al.}~\cite{ysoserial} have implemented the ysoserial tool 
which provides \totalNumberOfYsoserialPayloads{} publicly available Java deserialization payloads. 
These payloads represent gadget chains discovered in common Java libraries that can, under the right conditions, exploit Java applications performing unsafe deserialization of objects. When used, these payloads  lead to attacks that are critical since they allow, in most cases, the execution of arbitrary code. 
And even when attacks are unable to execute \emph{arbitrary} code, they may still be able to upload and delete files on the target host or send network traffic such as DNS requests. 
Arbitrary-code execution is the most common and most severe form of an attack since it is the first step  allowing the attacker to compromise the whole machine. 

In the same context, Haken has presented his tool \textit{Gadget Inspector}~\cite{gadgetinspector} to inspect Java libraries and classpaths for gadget chains. This tool allows to automatically detect possible gadgets chains in an application's classpath. Given war or jar file(s) of a library as input, this tool will go through several stages of classpath inspection to build up a list of blocks of gadget chains.  These blocks may exist in the full gadget chains discovered by ysoserial. Exploring the serialized byte streams to deserialize is relevant and may help to detect and locate some gadgets' information. In this context,  Bloor has developed a  tool named \emph{SerializationDumper}~\cite{SerializationDumper} which aims at automating the task of  decoding raw serialization streams and allows, then, to rebuild Java serialization streams and Java RMI packet contents in a more human-readable format. Unfortunately, this tool has some limitations like its inability to deserialize all Java serialized data streams and its "rebuild" mode which only operates on the Hex-Ascii encoded bytes from the dumped data. The same developer has provided another tool called \emph{DeserLab}~\cite{DeserLab}, a Java client and server application that implements a custom network protocol using the Java serialization format to demonstrate Java deserialization vulnerabilities.

Rasheed \textit{et al.} \cite{RasheedD20} propose a hybrid approach
that extends a static analysis with fuzzing to detect serialization
vulnerabilities. They use a heap abstraction to direct fuzzing for vulnerabilities in Java libraries.
Fingann~\cite{Fingann2020} presents  an overview of Java deserialization vulnerabilities, different techniques an attacker can use to exploit these vulnerabilities as well as which mitigation strategies can be employed to minimize the attack surface. 
This thesis encompasses the description of the most known works in the state of the art around deserializations vulnerabilities in Java: the Waratek description of deserialization problem~\cite{WaratekDeserProblem}, the tools like ysoserial~\cite{ysoserial} and Gadget Inspector~\cite{gadgetinspector}, the deserialization gadgets and chains of gadgets~\cite{RuntimeGadget}, and  the code practices to prevent deserialization attacks~\citep{PreventDeserVuln, JavaDeserializationSecurityFAQ}.  

Combating the deserialization attacks is one of the ideas explored by Seacord who examines in~\cite{seacord2017combating} Java deserialization vulnerabilities and evaluates various look-ahead object input streams solutions. 
Cristalli \textit{et al.}~\cite{CristalliVBL18} propose a sandboxing approach  for protecting Java applications based on a trusted execution path used for defining the deserialization behavior. 
They test their defensive mechanism on two Java frameworks, JBoss and Jenkins. They  design a sandbox system that is able to
intercept native methods by modifying the JVM internals.  Their sandboxing system performs two high-level phases: (1) dynamically analyzing Java applications and extracting the precise execution path in terms of stack traces, and
(2) use of a sandbox policy for monitoring applications at runtime and blocking incoming attacks: "\textit{when a native method is invoked by the application, the system intercepts it and checks whether the entire stack trace executed has been already observed in the learning phase. For this check, the system maintains a memory structure in the form of a hash table}"~\cite{CristalliVBL18}. \\

Deserialization vulnerabilities are not limited to the Java language. 
Shahriar \textit{et al.}~\cite{ShahriarH16} propose an approach to discover Object Injection Vulnerability (OIV) in PHP web applications. This kind of vulnerabilities involves accepting external inputs during deserialization operation. They use the concept of Latent Semantic Indexing (LSI~\cite{0021593}) to identify OIVs. Their approach was evaluated using three open-source PHP applications and was able to find the known OIV and to discover new vulnerabilities.

Dietrich \textit{et al.}~\cite{dietrich2017} study  serialization-related vulnerabilities for Java that exploit the topology of object graphs constructed from classes of the standard library. The deserialization, in this case, leads to resource exhaustion and denial of service
attacks. They analyze three vulnerabilities that can be exploited to exhaust stack memory, heap memory, and CPU time. They identify the language and library design features that enable these vulnerabilities. They demonstrate that these  Java vulnerabilities may  concern also C\#, JavaScript, and Ruby.

Peles \textit{et al.}~\cite{Peles2015} present high severity vulnerabilities in Android. One of these vulnerabilities concerns  the Android Platform and Google Play Services and allows arbitrary code execution. They perform a large-scale experiment over 32 701 Android applications and find new 
deserialization vulnerabilities unknown before. They demonstrate the impact of  the detected  vulnerabilities by developing a proof of concept exploit running under Google Nexus 5 Hammerhead running Android 5.1.1.

Alexopoulos \textit{et al.}~\cite{AlexopoulosHSM20}  presented a detailed analysis of the large body of open-source software packaged in the popular Debian GNU/Linux distribution. 
Vasquez \textit{et al.}~\cite{VasquezBE17} provided a study of Android-related vulnerabilities focusing on the ones affecting the Android OS. They have studied and classified 660 vulnerabilities. They have classified the  deserialization of untrusted data in a category called  "Indicator of poor quality code". 
The same goal was tackled by Mazuera \textit{et al.}~\cite{Mazuera-RozoBLR19} who  presented a large study aiming at analyzing software vulnerabilities in the Android OS. They analyzed 1235 vulnerabilities from different perspectives: vulnerability types and their evolution, CVSS vectors that describe the vulnerabilities, impacted Android OS layers, and their survivability across the Android OS history.

Other approaches against deserialization attacks suggest to use alternative data formats like textual ones (XML and JSON for example). The deserialization process using these formats does not invoke calls to gadgets, but, Fingann and  Muñoz et al.\citep{blackhatxml,Fingann2020} show that attackers may be able to perform their attacks regardless of the malicious data format.
XMLEncoder/XMLDecoder~\cite{xmlencodecodoc} is one of the existing mechanisms for using  alternative data formats like textual ones.
 While the XMLEncoder class is assigned to write output files for textual representation of Serializable objects, the XMLDecoder class reads an XML document that was created with XMLEncoder~\cite{redhatdeser}. The use of these alternative serialization mechanisms is not necessarily an effective solution to prevent  deserialization attacks. In fact,  Fingann ~\cite{Fingann2020} states that if an application uses XMLDecoder to deserialize a user byte stream, then the user may find another way to inject arbitrary code into the methods to call when deserializing the byte stream.  This means that any application  that uses  user input data to  deserialize by XMLDecoder  can be a victim of deserialization attacks.

Kyro~{\cite{kyro}} is an alternative Java implementation for serialization.
Similarly to the native Java implementation~{\cite{riggs1996pickling}}, it can lead to 
arbitrary code execution~{\cite{nist-CVE-2020-5413}}.

ProtocolBuffers~{\cite{protobufs}} is a generic approach to serialize any structured data.
It is available in many programming languages such as Java, Python, or C++.
Applications leveraging this serialization protocol might become vulnerable if they manipulate 
sensitive types such as {\texttt{java.lang.reflect.Method}}.
On top of that, the ProtocolBuffers' Java implementation itself was vulnerable to a 
denial of service attack~{\cite{nist-CVE-2021-22569}}.

Java deserialization filtering is the technique supported by Oracle~\cite{JavaDeserFilter}. 
Filters can validate incoming classes before they are deserialized by screening the  incoming streams of serialized objects.
For each new object in the stream that will be deserialized, the filters are invoked.
Support for serialization filters is included in Java 6 update 141, Java 7 update 131, Java 8 update 121, and all versions after Java 9.
Again, this technique requires the developer to manually extract serializable classes and add patterns to configure and activate the filter. \\ 

\subsection{Java Security}

Balzarotti \textit{et al.}~\cite{BalzarottiCFJKKV08} present an approach that  combines static and dynamic analysis techniques to identify faulty sanitization procedures that can be bypassed by an attacker through sensitive sinks in applications. The authors validate their approach by implementing the \textit{Saner} tool aiming at  analyzing the use of custom sanitization routines to identify possible XSS and SQL injection
vulnerabilities in web applications. They applied it to five real-world applications in  which they identify 13 vulnerabilities: for each sink, there exists at least one program path such that the output of a sanitization routine flows into this sink.
A systematic in-depth study of 87 publicly available Java exploits was performed in~\cite{HolzingerTBB16}. These attacks lead to security vulnerabilities that involve issues such as type confusion, deserialization issues, trusted method chaining, or confused deputies. Holzinger \textit{et al.}~\cite{HolzingerTBB16} show that all attack vectors implemented by the exploits belong to one of three categories: single-step attacks, restricted-class attacks, and information hiding attacks. They studied in detail the structures offered by the Java language such as the Security Manager feature and show how the analysis of the exploits samples helps for the detection of vulnerabilities.

Safe development of applications in Java relies a lot on the robustness of the JVM on which the code will be compiled and executed. 
Yuting \textit{et al.}~\cite{ChenSS19} dealt with the problem of validating the production of JVMs. 
The \textit{classfuzz} fuzzer allows the generation of illegal bytecode files that aim to test a JVM and  detect bugs in its bytecode verifiers. They developed their approach named \textit{classming} in the same context of validating JVMs. Their method is based on a technique called \textit{live bytecode mutation} able to generate mutant bytecode files from a seed bytecode file to test JVMs. They tested their approach on several  JVM implementations and reported the detected JVM crashes. One of their discoveries touches on a highly critical security vulnerability in Java 9 that allowed untrusted code to disable the Security Manager and elevate its privileges.
Confuzzion~\cite{2021confuzzion} is another JVM fuzzer which is more generic in the sense that it allows the generation of programs that are not possible to generate with Classming or Classfuzz.
Dean \textit{et al.}~\cite{DeanFW96} demonstrated that there is a significant number of flaws in the Java language and in these two browsers supporting it. They evoked the compromise that exists between the openness  desired  by  Web application writers and the security needs of their users. Their study aimed at finding the source of the identified flaws. They showed that the difference between the Java language and the bytecode semantics is one of the main reasons for weaknesses affecting the applications. A deeper study  of the bytecode and  bugs concluded that the Java system needs to be reviewed: the bytecode format and the runtime system should be redesigned in order to build a more secure system.

Holzinger \textit{et al.}~\cite{holzinger2017hardening} conducted a tool-assisted adaptation of the Java Class Library (JCL) able to  significantly harden the JCL against attacks. They study the problem of using shortcuts, originally introduced for ease of use and to improve performance, that cause Java to elevate the privileges of code implicitly. These shortcuts are responsible for a group of vulnerabilities known to have been exploited for the Java runtime: they directly enable attack vectors and complicate the security-preserving maintenance and evolution of the codebase by elevating privileges to certain callers implicitly. Their approach consists of three steps: (1) locate all shortcuts; (2) remove the shortcuts found and (3) wrap the calls in the JCL to those methods that formerly implemented shortcuts into privileged actions. 
Bartel \textit{et al.}~\cite{BartelKT19} presented an approach  based on a runtime solution called \textit{MUSTI}. This tool detects and prevents  invalid object initialization attacks. To achieve this goal, the authors patch the JVM by instrumenting the generated bytecode with an added code. This code checks if the objects have been correctly initialized. Their approach is generic and can be implemented in many languages supporting a similar sandbox system as the JVM.
 Any native code in Java programs may bypass the memory protection and the higher-level policies. To deal with these problems, Chisnall \textit{et al.}~\cite{chisnall2017cheri} developed a hardware-assisted implementation of the Java Native Interface (JNI), called CHERI JNI. This tool extends the guarantees required for Java’s security model to native code. Their approach  ensures  safe direct access to buffers owned by the JVM. 
%-------------------------------------------------------------------------------

%-------------------------------------------------------------------------------
\section{Conclusion}
\label{sec:conclusion}

The Java language is one of the most used languages to develop applications. 
Thanks to its ease of use and its portability, millions of applications run using this language. 
However, every year, many vulnerabilities in the Java runtime and its runtime libraries are discovered, reported, and patched. 
Our work highlights that vulnerabilities such as Java deserialization vulnerabilities can be critical since they impact the security of the applications, and in most cases allow the execution of arbitrary code.
In this paper, we have performed \numberOfExperiments ~experiments on \NumberStudiedAttacks ~RCE deserialization attacks.  
We have identified that not only the mentioned library versions in ysoserial attacks contain gadgets, but that there are previous and later versions that contain these gadgets as well. As an example, it is mentioned in this repository that the version 1.9.2 of the commons-beanutils library includes gadgets. After running our experiments, we found that 14 more versions contain also the same gadgets. These versions belong to the range between 1.5 and 1.9.4.  We have studied how gadgets are introduced in libraries and observe that the modification of one innocent-looking detail in a class - such as making it public - can already introduce a gadget.
Defense mechanisms such as filtering and allow/deny listing might be effective in preventing such attacks but might be hard to set up and maintain. 
We have performed an analysis of \JavaDeserCVEs ~CVEs - associated with vulnerable Java applications - from the Mitre database. We discovered that the results of our search on Mitre contain noise that we need to remove: among these \JavaDeserCVEs ~CVEs, only \DVCVEs ~CVEs represent "real" deserialization vulnerabilities description. The remaining 9 CVEs represent other types that we classified into GA (gadgets description) and UC (untrusted code description). 
We find that some patches of these application-level vulnerabilities consist in upgrading  library dependencies although these are frequently insufficient to prevent the deserialization attacks. Among the \DVCVEs ~studied CVEs, we were able to analyze \SuccessfulCVES ~CVEs in which 19\% are correctly fixed by disabling the deserialization of untrusted data.

A perspective for future work is the development of an algorithm for automatically detecting deserialization gadgets chains in applications, evaluate this implementation on a set of applications and  compare it with other existing tools such as Gadget Inspector~\cite{gadgetinspector}. Our goal is to prevent deserialization attacks early.  
Identifying the parameters involved to define the duration of patching libraries  is another perspective of this work. In this paper, we have focused only on RCE attacks since we have designed a framework  to test only this kind of attacks. Now, we are planning an extension of this framework to cover the remaning 15 non-RCE ysoserial attacks containing DOS ones which aim at making services unavailable to their legitimate users. 
%-------------------------------------------------------------------------------

\section*{Acknowledgments}
This work was supported by the Luxembourg National Research Fund (FNR) ONNIVA
Project, ref. 12696663.
This work was partially supported by
the Wallenberg AI, Autonomous Systems and Software Program (WASP) funded by the
Knut and Alice Wallenberg Foundation.

\bibliographystyle{splncs04} % ACM-Reference-Format 
 \bibliography{bib}

\begin{appendices}

\newpage
\begin{landscape}
\section{Vulnerable applications and their patches}
\label{apps-vulnerable}

\footnotesize {
\begin{longtable}{|p{0.13\textwidth}|p{0.1\textwidth}|p{0.26\textwidth}|p{0.31\textwidth}|p{0.4\textwidth}|p{0.025\textwidth}|p{0.025\textwidth}|p{0.025\textwidth}|}
\hline
\textbf{Application} & \textbf{CVE} & \textbf{Code availability} & \textbf{Vulnerability description} & \textbf{Applied patch} & \textbf{GA} & \textbf{DV} & \textbf{UC} \\
\hline
WebSphere Application Server (WAS) Community Edition 3.0.0.3 & CVE-2013-1777 & \url{http://geronimo.apache.org/downloads.html} & Remote exploits can be prevented by hiding the naming (1099) and JMX (9999) ports behind a firewall or binding the ports to a local network interface. & Add instruction \texttt{ Thread.currentThread().set\-ContextClassLoader(get\-Class().getClassLoader());} in class \texttt{JMXConnector} and other instructions in the class \texttt{JMXSecureConnector} (patch in \url{http://svn.apache.org/viewvc?view=revision&sortby=date&revision=1458113}) & & X & \\
\hline
Jboss RichFaces (Jboss-RF) 3.x $<=$ 3.3.3 and 4.x $<=$ 4.3.2 & CVE-2013-2165 & \url{https://richfaces.jboss.org/download/archive.html} & A flaw in the way \texttt{JBoss RichFaces} handled deserialization allowing a remote attacker to trigger the execution of the deserialization methods in any serializable class deployed on the server. & Create a whitelist of classes that are available to participate in the \texttt{RichFaces} resource deserialisation process \url{https://www.bleathem.ca/blog/richfaces-security-advisory-cve-2013-2165/} and \url{https://codewhitesec.blogspot.com/2018/05/poor-richfaces.html}  & & X & \\
\hline
Android $<$ 5.0.0 & CVE-2014-7911 & \href{https://android.googlesource.com/?format=HTML}{https://android.googlesour\-ce.com/?format=HTML} & \texttt{luni/src/main/java/java/io/\-ObjectInputStream.java} in the \texttt{java.io.ObjectInputStream} implementation does not verify that deserialization will result in an object that met the requirements for serialization, which allows attackers to execute arbitrary code via a crafted \texttt{finalize} method for a serialized object in an \texttt{ArrayMap} Parcel within an intent sent to system\_service, as demonstrated by the \texttt{finalize} method of \texttt{android.os.BinderProxy} & Add some checks that the class being deserialized matches the type information (\texttt{enum}, \texttt{serializable}, \texttt{externalizable}) held in the stream. Delayed static initialization of classes until the type of the class has been validated against the stream content in some cases. (see \href{https://android.googlesource.com/platform/libcore/+/738c833d38d41f8f76eb7e77ab39add82b1ae1e2\%5E\%21/#F0}{https://android.googlesource.com/platform/\-libcore/+/738c833d38d41f8f76eb7e77ab39a\-dd82b1ae1e2\%5E\%21/\#F0} and \href{https://android.googlesource.com/platform/libcore/+/738c833d38d41f8f76eb7e77ab39add82b1ae1e2}{https://android.googlesource.com/platform/\-libcore/+/738c833d38d41f8f76eb7e77ab39a\-dd82b1ae1e2}) &  & X &  \\
\hline
Atlassian Bamboo before 5.9.9 and 5.10.x before 5.10.0 & CVE-2014-9757 & \url{https://www.atlassian.com/software/bamboo/download-archives} & The Ignite Realtime Smack XMPP API, as used in Atlassian Bamboo before 5.9.9 and 5.10.x before 5.10.0, allows remote configured XMPP servers to execute arbitrary Java code via serialized data in an XMPP message & The origin of the attack is the Smack library used in Bamboo. The patched version Bamboo 5.10.0 uses an updated version of the smack library in which a lot of modifications are brought: removing some classes (like \texttt{Connection}, \texttt{Chat}, \texttt{ConnectionManager}), modify the class \texttt{XMPPConnection} into an \texttt{Interface} , etc. (Patch obtained doing the diff between the version 5.9.7 and 5.10.0 of Bamboo, and more preciely the smack library).  & & X & \\
\hline
Atlassian Bamboo 2.2 before 5.8.5 and 5.9.x before 5.9.7 & CVE-2015-6576 & \url{https://www.atlassian.com/software/bamboo/download-archives} & Bamboo 2.2 before 5.8.5 and 5.9.x before 5.9.7 allows remote attackers with access to the Bamboo web interface to execute arbitrary Java code via an unspecified resource. & Removes the \texttt{deserializeObject} method from the \texttt{DeliverMessageServlet} vulnerable class. & & X & \\
\hline
Atlassian Bamboo before 5.9.9 and 5.10.x before 5.10.0 & CVE-2015-8360 & \url{https://www.atlassian.com/software/bamboo/download-archives} & An unspecified resource in Atlassian Bamboo before 5.9.9 and 5.10.x before 5.10.0 allows remote attackers to execute arbitrary Java code via serialized data to the JMS port. & Use of black and white lists for serialization (patch obtained using the diff between the versions 5.10.0 and 5.9.7: there are two files \texttt{serialization-blacklist.list} and \texttt{serialization-whitelist.list} in the path "atlassian-bamboo-5.10.0/atlassian-bamboo/WEB-INF/classes") & & X & \\
\hline
Jenkins $<$ 1.638 and LTS $<$ 1.625.2 & CVE-2015-8103 & \url{https://github.com/jenkinsci/jenkins} \url{https://wiki.jenkins.io/display/JENKINS/Jenkins+CLI} & The Jenkins CLI subsystem in Jenkins before 1.638 and LTS before 1.625.2 allows remote attackers to execute arbitrary code via a crafted serialized Java object  & Unknown patch. Mitigation: remove/disable the CLI support inside of the running Jenkins server (\url{https://www.jenkins.io/blog/2015/11/06/mitigating-unauthenticated-remote-code-execution-0-day-in-jenkins-cli/}) & & X & \\
\hline
VMware vRealize Orchestrator 6.x, vCenter Orchestrator 5.x, vRealize Operations 6.x, vCenter Operations 5.x, and vCenter Application Discovery Manager (vADM) 7.x & CVE-2015-6934 & \url{https://docs.vmware.com/en/vRealize-Orchestrator/7.6/rn/VMware-vRealize-Orchestrator-76-Release-Notes.html} & Remote attackers can execute arbitrary commands via a crafted serialized Java object, related to the Apache Commons Collections library & Replace the Commons collections library by the commons-collections-3.2.2.jar in the dependencies of the mentioned products (see \url{https://kb.vmware.com/s/article/2141244} and \url{https://kb.vmware.com/s/article/2141244}) &  & X &  \\
\hline
Adobe Experience Manager (Adobe-EM) 5.6.1, 6.0.0, and 6.1.0 & CVE-2016-0958 & No & Adobe Experience Manager 5.6.1, 6.0.0, and 6.1.0 might allow remote attackers to have an unspecified impact via a crafted serialized Java object. & Unknown patch & & X &  \\
\hline
Hazelcast $<$ 3.11 & CVE-2016-10750 & \url{https://github.com/hazelcast/hazelcast} & A flaw was found in the cluster join procedure in Hazelcast. This flaw allows an attacker to gain remote code execution via Java deserialization. & Add class names blacklisting and whitelisting by defining  the following system properties: \texttt{hazelcast.serialization.filter.enabled},  \texttt{hazelcast.serialization.filter.black\-list.classes}, \texttt{hazelcast.serialization.filter.black\-list.packages}, \texttt{hazelcast.serialization.filter.white\-list.classes} and \texttt{hazelcast.serialization.filter.white\-list.packages} (see \url{https://docs.hazelcast.org/docs/3.10.5/manual/html-single/index.html#untrusted-deserialization-protection}) &  & X &  \\
\hline
Apache OFBiz 12.04.x $<$ 12.04.06 and 13.07.x $<$ 13.07.03 & CVE-2016-2170 & \url{http://archive.apache.org/dist/ofbiz/} &  Remote attackers can execute arbitrary commands via a crafted serialized Java object, related to the Apache Commons Collections library & Update commons collections to 4.1 and Comment out RMI related code (see \url{https://issues.apache.org/jira/browse/OFBIZ-6942},  \url{https://markmail.org/message/nh6csf4fun5n6e23} and \url{https://issues.apache.org/jira/browse/OFBIZ-6726}) &  & X &  \\
\hline
SolarWinds Virtualization Manager $<=$ 6.3.1 & CVE-2016-3642 & No & The vulnerability exists due to the deserialization of untrusted data in the RMI service running on port 1099/TCP. A remote attacker can execute operating system commands as an unprivileged user\footnote{information about a cyber attack in Solarwinds are available in \url{https://www.secureworld.io/industry-news/solarwinds-cyber-attack-impact-update}} &  Inaccessible patch (it is mentioned that there is a hotfix in \url{https://packetstormsecurity.com/files/137486/Solarwinds-Virtualization-Manager-6.3.1-Java-Deserialization.html} and \url{https://seclists.org/fulldisclosure/2016/Jun/29} but no more details are given) &  & X &  \\
\hline
HP Network Node Manager i (HP-NNMi) Software 10.00, 10.01 (patch1), 10.01 (patch 2), 10.10  & CVE-2016-4398 & No & A vulnerability in Apache Commons Collections for handling Java object deserialization was addressed by HPE Network Node Manager i (NNMi) Software. The vulnerability could be remotely exploited to allow remote code execution. & Unknown patch &  & X &  \\
\hline
Apache Wicket 6.x $<$ 6.25.0 and 1.5.x $<$ 1.5.17 & CVE-2016-6793 & \url{https://archive.apache.org/dist/wicket/} & The \texttt{DiskFileItem} class in Apache Wicket allows remote attackers to cause a denial of service (infinite loop) and write to, move, and delete files with the permissions of \texttt{DiskFileItem}, and if running on a Java VM before 1.3.1, execute arbitrary code via a crafted serialized Java object. & Change the class \texttt{DiskFileItem}: add a check instruction \texttt{Files.checkFileName(tempDir.getPath())} in the method \texttt{getTempFile()} of the patched version (patch obtained by doing the diff betwwen the 6.24.0 and 6.25.0 versions) &  & X &  \\
\hline
Red Hat JBoss Enterprise Application Platform (Jboss-EAP) 4 and 5 & CVE-2016-7065 & \url{https://developers.redhat.com/products/eap/download}  & JBoss EAP 4 and 5 JMX servlet is exposed on port 8080/TCP with authentication by default. The communication employs serialized Java objects, encapsulated in HTTP requests and responses. The server deserializes these objects. This behavior can be exploited to cause a denial of service and potentially execute arbitrary code & Red Hat does not fix the issue because JBoss EAP 4 is out of maintenance
support and JBoss EAP 5 is close to the end of its maintenance period (see \url{https://seclists.org/fulldisclosure/2016/Nov/143} and \url{https://seclists.org/fulldisclosure/2016/Nov/143}) &  & X &  \\
\hline
Soffid IAM $<$ 1.7.5 & CVE-2017-9363 & \url{https://github.com/SoffidIAM/console} & Untrusted Java serialization in Soffid IAM console before 1.7.5 allowing remote attackers to achieve arbitrary remote code execution via a crafted authentication request  & Disable two features in the class  \texttt{servlet.SignatureReceiver} via throwing two exceptions \texttt{new ServletException("Disabled feature");} and \texttt{new UiException("Disabled feature");} (see \href{https://github.com/SoffidIAM/console/commit/8e9e7c9e537acfc2a245fbbeb41a143b5b4f7230#diff-544c1cb1ac64f2f62b6b326bd0b1b6addc17f19416878d319d3643e302a043b7}{https://github.com/SoffidIAM/console/com\-mit/8e9e7c9e537acfc2a245fbbeb41a143b5b\-4f7230\#diff-544c1cb1ac64f2f62b6b326bd0b\-1b6addc17f19416878d319d3643e302a043b7}) & & X & \\
\hline
ZTE ZXIPTV-EPG $<$ V5.09.02.02T4 &  CVE-2017-10934 & No & This product uses the Java RMI service in which the servers use the Apache Commons Collections (ACC) library that may result in Java deserialization vulnerabilities. An unauthenticated remote attacker can exploit the vulnerabilities by sending a crafted RMI request to execute arbitrary code on the target host & Unknown patch. Workaround: Ensure that all exposed ports used by the server, including the RMI registry port, are firewalled from any untrusted IP address. (see \url{http://support.zte.com.cn/support/news/LoopholeInfoDetail.aspx?newsId=1008682}) &  & X &  \\
\hline
Akka versions $<=$ 2.4.16 and 2.5-M1 & CVE-2017-1000034 & \url{https://mvnrepository.com/artifact/com.typesafe.akka/akka-actor_2.12} & An attacker that can connect to an \texttt{ActorSystem} exposed via Akka Remote over TCP can gain remote code execution capabilities in the context of the JVM process that runs the \texttt{ActorSystem} under some conditions (\texttt{JavaSerializer} is enabled (default in Akka 2.4.x),  etc.) & The system is configured with disabled Java serializer: using \texttt{DisabledJavaSerializer} instead of \texttt{JavaSerializer} (see the file \texttt{reference.conf} for explanation). Additional protection can be achieved when running in an untrusted network by enabling TLS with mutual authentication. \url{https://doc.akka.io/docs/akka/2.4/security/2017-02-10-java-serialization.html}, \url{https://akka.io/blog/news/2017/02/10/akka-2.4.17-released}  and \url{https://doc.akka.io/docs/akka/2.4/scala/remoting.html#remote-tls-scala}&  & X &  \\
\hline
Cisco Unity Express (Cisco-UE) $<$ release 9.0.6  & CVE-2018-15381  & No &  A remote user can create specially crafted content that, when loaded by the target user, will trigger a Java deserialization flaw and execute arbitrary code on the target user's system. The code will run with root privileges. & Workaround: this vulnerability can be exploited over TCP port 1099. The CUE does not need this port to be open externally and may be blocked to protect against remote exploitation of this vulnerability. An administrator can configure an access control list that blocks all traffic with a destination port of TCP/1099 from reaching the CUE. (see \url{https://tools.cisco.com/security/center/content/CiscoSecurityAdvisory/cisco-sa-20181107-cue}) &  & X &  \\
\hline
Apache Storm versions 1.1.0 to 1.2.2 &  CVE-2018-11779 & \url{https://archive.apache.org/dist/storm/} & When the user is using the storm-kafka-client or storm-kafka modules, it is possible to cause the Storm UI daemon to deserialize user provided bytes into a Java class. & Update implementation of serializable classes in v1.2.3: remove the indirect call to \texttt{readObject} from \texttt{getSetComponentObject} method (using the diff between the vulnerable 1.2.2 version and the non vulnerable 1.2.3 version) &  & X &  \\
\hline
Jenkins Pipeline supporting APIs Plugin $<=$ 2.17 & CVE-2018-1000058 & \url{https://updates.jenkins.io/download/plugins/workflow-support/} & Methods related to Java deserialization like \texttt{readResolve} implemented in Pipeline scripts were not subject to sandbox protection, and could therefore execute arbitrary code. This could be exploited e.g. by regular Jenkins users with the permission to configure Pipelines in Jenkins, or by trusted committers to repositories containing \texttt{Jenkinsfiles}. & Adding sandboxing: reinforcement of the class \texttt{RiverWriter} using a \texttt{try/catch} bloc in which the serialization is carried out inside a \texttt{GroovySandbox.runInSandbox()} method. The class \texttt{RiverReader} is also strengthen by performing the deserialization inside a sandbox presented by an inner class \texttt{SandboxedUnmarshaller}. The patch s found using the diff between the 2.17 (vulnerable) and 2.18 (patched) versions. &  & X &  \\
\hline
Log4j  & CVE-2019-17571 & \url{https://github.com/apache/log4j} & A vulnerable \texttt{SocketServer} class may lead to the deserialization of untrusted data allowing an attacker to remotely execute arbitrary code when combined with a deserialization gadget & Add class filtering to \texttt{AbstractSocketServer}:
this allows a whitelist of class names to be specified to configure
which classes are allowed to be deserialized in both TcpSocketServer and
UdpSocketServer  (link: \url{https://git-wip-us.apache.org/repos/asf?p=logging-log4j2.git;h=5dcc192}) & & X & \\
\hline
JetBrains TeamCity before 2019.1.4 & CVE-2019-18364 & \url{https://www.jetbrains.com/fr-fr/teamcity/download/other.html} & Insecure Java Deserialization could potentially allow remote code execution & Unknown patch (researching patch exceeds time limit) &  & X &  \\
\hline
Apache Dubbo 2.7.0 before 2.7.5, 2.6.0 before 2.6.8, and 2.5.x versions & CVE-2019-17564 & \url{https://github.com/apache/dubbo} & An attacker may submit a POST request with a Java object in it to completely compromise a \texttt{Provider} instance of Apache Dubbo, if this instance enables HTTP. & The patched version does not support outdated \texttt{http-invoker rpc} protocol anymore   (see \url{https://github.com/apache/dubbo/commit/9b18fe228971eaeca9b87d7b7e95df1c2a8ff91b} and \url{https://github.com/apache/dubbo/releases/tag/dubbo-2.7.5}) &  & X &  \\
\hline
 Apache Ofbiz from 16.11.01 to 16.11.05  & CVE-2019-0189 & \url{https://archive.apache.org/dist/ofbiz/} & This issue  is exposed by the \texttt{"webtools/control/httpSer\-vice"} URL, and uses Java deserialization to perform code execution. In the \texttt{HttpEngine}, the value of the request parameter \texttt{serviceContext} is passed to the \texttt{deserialize}  method of \texttt{XmlSerializer}. & Improve \texttt{ObjectInputStream} class and redefine it as a new class \texttt{SafeObjectInputStream} in which there is an added  whitelist. Also add objects from \texttt{org.apache.commons.fileupload} (namely \texttt{DiskFileItem} and \texttt{FileItemHeadersImpl}) as non-serializable in this class \texttt{SafeObjectInputStream} (see the diff between the two versions 16.11.05 and 16.11.06. See also \href{https://gitbox.apache.org/repos/asf?p=ofbiz-framework.git;a=blob;f=framework/base/src/main/java/org/apache/ofbiz/base/util/SafeObjectInputStream.java;h=d50cfbf11fc4d3b5855c53cb38a6cde7e101dc83;hb=3f60efb}{https://gitbox.apache.org/repos/asf?p=of\-biz-framework.git;a=blob;f=framework/ba\-se/src/main/java/org/apache/ofbiz/base/\-util/SafeObjectInputStream.java;h=d50cf\-bf11fc4d3b5855c53cb38a6cde7e101dc83;h\-b=3f60efb}) &  & X &  \\
\hline
Apache Tapestry & CVE-2019-0195 & \url{https://downloads.apache.org/tapestry/}  & Manipulating classpath asset file URLs, an attacker could guess the path to a known file in the classpath and have it downloaded. It is possible to download arbitrary class files from the classpath by providing a crafted asset file URL. An attacker is able to download the file \texttt{AppModule.class} by requesting the URL \url{'http://localhost:8080/assets/something/services/AppModule.class'} which contains a HMAC secret key. & The fix for that bug was a blacklist filter that checks if the URL ends with \texttt{'.class'}, \texttt{'.properties'}  or \texttt{'.xml'}. However, it is proven that this blacklist solution can simply be bypassed by appending a '/' at the end of the URL: \url{'http://localhost:8080/assets/something/services/AppModule.class/'} (source: \href{https://lists.apache.org/thread.html/r237ff7f286bda31682c254550c1ebf92b0ec61329b32fbeb2d1c8751@\%3Cusers.tapestry.apache.org\%3E}{https://lists.apache.org/thread.html/r237ff\-7f286bda31682c254550c1ebf92b0ec61329b\-32fbeb2d1c8751@\%3Cusers.tapestry.apache.\-org\%3E})&  & X &  \\
\hline
Apache Tomcat & CVE-2020-9484 & \url{https://github.com/apache/tomcat} & Deserialization flaw in session persistence storage FileStore leading to remote code execution & Update the class \texttt{FileStore} with some checks (patch in \url{https://github.com/apache/tomcat/commit/bb33048e3f9b4f2b70e4da2e6c4e34ca89023b1b})  & & X & \\
\hline
OpenNMS Horizon $<$ 26.0.1 and Meridian before 2018.1.19 and 2019 before 2019.1.7 & CVE-2020-12760 & \url{https://github.com/OpenNMS/opennms/releases/tag/opennms-26.0.1-1} & The ActiveMQ channel configuration allowed for arbitrary deserialization of Java objects  leading to remote code execution for any authenticated channel user regardless of its assigned permissions & Remove a parameter after  stopping the use of serialized object messages in a file \texttt{applicationContext-daemon.xml}:  
 \texttt{<property name="trustAllPackages" value="true"/>} (see    \url{https://github.com/OpenNMS/opennms/pull/2983} and \url{https://github.com/OpenNMS/opennms/pull/2983/files/e21fc14ce355533493da0db815bd81a66e291382})  \url{https://github.com/davidhalter/parso/issues/75#}) & & X & \\
\hline
IBM Maximo Asset Management 7.6.0 and 7.6.1 & CVE-2020-4521 & \url{https://github.com/nishi2go/maximo-docker} & IBM Maximo Asset Management could allow a remote authenticated attacker to execute arbitrary code on the system, caused by an unsafe deserialization in Java. By sending specially-crafted request, an attacker could exploit this vulnerability to execute arbitrary code on the system & Inaccessible patch (when connecting to \url{https://www.ibm.com/support/pages/node/6332587} an error message ("No applicable IBM support agreement found for one or more of the products you selected") appears) &  & X &  \\
\hline
Cisco Security Manager (Cisco-SM) & CVE-2020-27131 & No & Multiple vulnerabilities in the Java deserialization function that is used by Cisco Security Manager could allow an unauthenticated, remote attacker to execute arbitrary commands on an affected device. These vulnerabilities are due to insecure deserialization of user-supplied content by the affected software. An attacker could exploit these vulnerabilities by sending a malicious serialized Java object to a specific listener on an affected system. A successful exploit could allow the attacker to execute arbitrary commands on the device with the privileges of NT AUTHORITY$\textbackslash$SYSTEM on the Windows target host. Cisco has not released software updates that address these vulnerabilities & Unknown patch &  & X &  \\
\hline
Taoensso Nippy $<$  2.14.2 & CVE-2020-24164 & \url{https://github.com/ptaoussanis/nippy} & A deserialization flaw is present in Taoensso Nippy before 2.14.2. In some circumstances, it is possible for an attacker to create a malicious payload that, when deserialized, will allow arbitrary code to be executed. This occurs because there is automatic use of the Java \texttt{Serializable} interface: Nippy introduced a feature to allow the automatic use of Java's \texttt{Serializable} interface as a fallback for types that Nippy didn't support via its own \texttt{Freezable} protocol. & Use a predicate \texttt{(fn allow-class? [class-name]) fn} that can be assigned to \texttt{'*freeze-serializable-allowlist*'} and/or \texttt{'*thaw-serializable-allowlist*'}. This predicate is used to record information about which classes have been using Nippy's \texttt{Serializable} support in the user's environment (see \url{http://ptaoussanis.github.io/nippy/taoensso.nippy.html#var-allow-and-record-any-serializable-class-unsafe}) &  & X &  \\
\hline
Apache Tapestry 4 & CVE-2020-17531 & \url{https://github.com/apache/tapestry4}  & Apache Tapestry 4 will attempt to deserialize the "sp" parameter even before invoking the page's validate method, leading to deserialization without authentication & Apache Tapestry 4 reached end of life in 2008 and no update to address this issue is released (the upgrade to the latest Apache Tapestry 5 version is necessary) (see \href{https://lists.apache.org/thread.html/r700a6aa234dbff0555d4187bdc8274d7e4c0afbf35b9a3457f09ee76\%40\%3Cusers.tapestry.apache.org\%3E}{https://lists.apache.org/thread.html/r700a\-6aa234dbff0555d4187bdc8274d7e4c0afbf35b9\-a3457f09ee76\%40\%3Cusers.tapestry.apache.\-org\%3E}) and \url{https://cve.mitre.org/cgi-bin/cvename.cgi?name=CVE-2020-17531}) & & X &  \\
\hline
Gradle Enterprise Maven Extension & CVE-2020-15777 & \url{https://mvnrepository.com/artifact/com.gradle/gradle-enterprise-maven-extension} & The extension uses a socket connection to send serialized Java objects. Deserialization is not restricted to an allow-list, thus allowing an attacker to achieve code execution via a malicious deserialization gadget chain. The socket is not bound exclusively to localhost. The port this socket is assigned to is randomly selected and is not intentionally exposed to the public (either by design or documentation). This could potentially be used to achieve remote code execution and local privilege escalation. & Add an allow-list in a class \texttt{ValidatingObjectInputStream} (patch obtained by doing the diff between the vulnerable 1.5.3 and the non-vulnerable 1.6 versions) &  & X &  \\
\hline
Apache Camel Netty (Camel-Netty) & CVE-2020-11973 & \url{https://github.com/apache/camel/tree/main/components/camel-netty} & Apache Camel RabbitMQ enables java deserialization, by default, without any means of disabling which can lead to arbitrary code being executed. The highest threat from this vulnerability is to data confidentiality and integrity as well as system availability & Disable object serialization: only Strings are allowed to be serialized by default, anything else will only be serialized with a custom encoder/decoder (\url{https://github.com/apache/camel/pull/3537})  &  & X &  \\
\hline
Apache Camel RabbitMQ (Camel-RabbitMQ) 2.22.x, 2.23.x, 2.24.x, 2.25.0, 3.0.0 up to 3.1.0  & CVE-2020-11972 & \url{https://github.com/apache/camel} & Apache Camel RabbitMQ enables java deserialization, by default, without any means of disabling which can lead to arbitrary code being executed. The highest threat from this vulnerability is to data confidentiality and integrity as well as system availability & Disable RabbitMQ Java serialization by default. It can be re-enabled using a parameter \texttt{"allowMessageBodySerialization"} in a class \texttt{RabbitMQEndpoint} (see \url{https://github.com/zregvart/camel/commit/c15ed20d92b5c920e9e55fe584f8e412b23f14f6}) &  & X &  \\
\hline
Emissary 6.4.0 & CVE-2021-32634 & \url{https://github.com/NationalSecurityAgency/emissary} & Unsafe Deserialization of post-authenticated requests to the WorkSpaceClientEnqueue.action REST endpoint. & Remove unsafe serialization from \texttt{PayloadUtil}. Remove the class \texttt{WorkBundle} from the list of serializable classes, remove some classes like \texttt{MoveToAction} and \texttt{MoveToAdapter}. Replace the \texttt{ObjectInputStream} by \texttt{DataInputStream} (\url{https://github.com/NationalSecurityAgency/emissary/security/advisories/GHSA-m5qf-gfmp-7638} & & X & \\
\hline
Apache Dubbo prior to 2.6.9 and 2.7.9 & CVE-2021-30179 & \url{https://github.com/apache/dubbo} & Apache Dubbo by default supports generic calls to arbitrary methods exposed by provider interfaces.
These invocations are handled by the \texttt{GenericFilter} which will find the service and method specified in the first arguments of the invocation and use the Java Reflection API to make the final call. The signature for the \texttt{invoke} or \texttt{invokeAsync} methods is \texttt{Ljava/lang/String;\-[Ljava/lang/String;[Ljava/lang\-/Object;} where the first argument is the name of the method to invoke, the second one is an array with the parameter types for the method being invoked and the third one is an array with the actual call arguments & Native Java deserialization will not be activated defaultly. If user still wants use it,	 set \texttt{ dubbo.security.serialize.generic.nati\-ve-java-enable} as \texttt{true} in environment. An embedded serialization block list is introduced in \texttt{dubbo-common/src/main/resources/secu\-rity/serialize.blockedlist}.  (see \url{https://github.com/apache/dubbo/releases/tag/dubbo-2.7.10}) & & X & \\
\hline
\multirow{2}{0.13cm}{Apache OFBiz} & CVE-2021-29200  & \url{https://github.com/apache/ofbiz-framework} & An unauthenticated user can perform an RCE attack  & Update \texttt{UtilObject} class. Restrict unauthorized deserialisations to \texttt{java.rmi} instead of \texttt{java.rmi.server}. (patch in \url{https://issues.apache.org/jira/browse/OFBIZ-12216} and \url{https://gitbox.apache.org/repos/asf?p=ofbiz-framework.git;h=1bc8a20}) &  & X &  \\
\cline{2-8}
& CVE-2021-26295 & \url{https://github.com/apache/ofbiz-framework} & An unauthenticated attacker can use this vulnerability to successfully take over Apache OFBiz & The code fix is to "blacklist" RMI server to prevent it from being exploited. (see \url{https://issues.apache.org/jira/browse/OFBIZ-12167} and \href{https://lists.apache.org/thread.html/r0d97a3b7a14777b9e9e085b483629d2774343c4723236d1c73f43ff0@\%3Cdev.ofbiz.apache.org\%3E}{https://lists.apache.org/thread.html/r0d97\-a3b7a14777b9e9e085b483629d2774343c4723\-236d1c73f43ff0@\%3Cdev.ofbiz.apache.org\%3E} &  & X &  \\
\hline
\multirow{2}{0.2cm}{McAfee Database Security (DBSec) $<$ 4.8.2} & CVE-2021-23895 & No & A remote authenticated attacker can  create a reverse shell with administrator privileges on the DBSec server via carefully constructed Java serialized object sent to the DBSec server & Unknown patch &  & X &  \\
\cline{2-8}
 & CVE-2021-23894  & No & A remote unauthenticated attacker can create a reverse shell with administrator privileges on the DBSec server via carefully constructed Java serialized object sent to the DBSec server. & Unknown patch &  & X &  \\
\hline
\caption{29 studied CVEs and applied patches. The first column designates the name of the studied vulnerable application; the column "CVE" mentions the CVE ID associated to the vulnerability; the third column "Code availability" indicates if the source code or the binary files are available: if yes, we give the URL for this code, otherwise we put "No"; the description of each vulnerability is presented in the fourth column; patching or workaround actions are described in the column "Applied patch"; the last three columns desingate the category of the vulnerability at hand: GA for GAdgets, DV for Deserialization Vulnerabilities and UC for Untrusted Code. The rows  having UC as category are colored in gray because they are not in our scope of study in this article. Note that the complete table, with \JavaDeserCVEs~ CVEs, is available at \url{https://github.com/software-engineering-and-security/java-deserialization-rce}}
  \label{tab:app-vuln}
\end{longtable}
} 
\end{landscape}

\begin{landscape}
\section{Pre-processing for use of filters} 
\label{type-extract}

\begin{longtable}{p{0.25\textwidth}p{0.6\textwidth}p{0.35\textwidth}}
\hline
\textbf{Attack name} & \textbf{(De)Serialized types} & \textbf{Is the first object to deserialize in external library type (Y/N)?} \\
\hline
\emph{BeanShell1} & PriorityQueue$<$E$>$ (containing an int and an XThis\$Handler), E is an Integer, This, BshMethod & N \\
\hline
\emph{Clojure} &  HashMap$<$K,V$>$ (containing int and float), K is an AbstractTableModel\$ff19274a &  N \\
\hline
\emph{CommonsBeanUtils1} & PriorityQueue$<$E$>$ (containing an int and a BeanComparator$<$T$>$ (in which T is a TemplatesImpl), E is a TemplatesImpl & N \\
\hline
\emph{CommonsCollections1} & AnnotationInvocationHandler (containing Class$<$? extends Annotation$>$, Map$<$String, Object$>$ which is a LazyMap containing ChainedTransformer and a HashMap)), InvokerTransformer & N \\
 \hline
 \emph{CommonsCollections2} & PriorityQueue$<$E$>$ (containing int, TransformingComparator), E is a TemplatesImpl, InvokerTransformer & N \\
 \hline
 \emph{CommonsCollections3} & AnnotationInvocationHandler, InstantiateTransformer, TemplatesImpl & N \\
 \hline
 \emph{CommonsCollections4} & PriorityQueue$<$E$>$ (containing int, TransformingComparator), E is an Integer, ChainedTransformer, InstantiateTransformer and TemplatesImpl & N \\
 \hline
 \emph{CommonsCollections5} & BadAttributeValueExpException (containing Object (which is a String)), TiedMapEntry, LazyMap, ChainedTransformer, InvokerTransformer & N \\
 \hline
 \emph{CommonsCollections6} & HashSet$<$E$>$ (containing an Object), E is a TiedMapEntry, LazyMap, ChainedTransformer, InvokerTransformer &  \\
 \hline
 \emph{CommonsCollections7} & HashTable$<$K,V$>$ (containing int and float), K is a LazyMap and V is an Integer, ChainedTransformer, InvokerTransformer  & N \\
 \hline
\emph{Groovy1} & AnnotationInvocationHandler (containing a ConvertedClosure and Class$<$T$>$), ConversionHandler, Closure$<$V$>$ &  N \\
 \hline
\emph{ROME} & HashMap$<$K,V$>$ (containing int, float, ObjectBean), EqualsBean, ToStringBean, TemplatesImpl & N \\
 \hline
 \emph{MozillaRhino1} & BadAttributeValueExpException (containing an Object which is a String), ScriptableObject, TemplatesImpl & N \\
 \hline
 \emph{MozillaRhino2} & NativeJavaObject (containing org.mozilla.javascript.tools.shell.Environment, Scriptable interface), ScriptableObject, NativeJavaObject, MemberBox, TemplatesImpl & Y (class in js-1.7-R2) \\
 \hline
\emph{Spring1} & SerializableTypeWrapper\$MethodInvokeTypeProvider (containing TypeProvider (which is a TemplatesImpl), String and int) & Y (in spring-core library) \\ 
 \hline
 \emph{Spring2} & SerializableTypeWrapper\$MethodInvokeTypeProvider (containing String, int and TypeProvider (which is a TemplatesImpl)), JdkDynamicAopProxy & Y (in spring-core library) \\ 
 \hline
 \emph{Click1} & PriorityQueue$<$E$>$ (containing int, Column\$ColumnComparator), E is a TemplatesImpl, Column & N \\ 
 \hline
 \emph{Vaadin1} & BadAttributeValueExpException (containing an Object which is a String), TemplatesImpl & N \\
 \hline
  \emph{JDK7U21} & LinkedHashSet$<$E$>$ (HashSet$<$E$>$) (containing Object), E is a TemplatesImpl, HashMap$<$K,V$>$, AnnotationInvocationHandler  &  N \\
\hline
\caption{Types extracted from known attacks} 
\label{tab:type-extract}
\end{longtable}

\newpage

\begin{longtable}{p{0.25\textwidth}p{0.8\textwidth}}
\hline
\textbf{Application name} & \textbf{Deserialized types} \\  
\hline
\emph{Apache-wicket (6.24.0)} &  String, ObjectInputStream,  ReplaceObjectInputStream (containing HashMap$<$String, Component$>$ (), ClassLoader (containing boolean, Hashtable, Certificate[], Vector, HashMap, Set, URLClassPath)), List$<$Serializable$>$, Map$<$String, SessionEntry$>$ (for SessionEntry, there are String, boolean, PageWindowManager (containing  PageWindowInternal (with int, long, List$<$PageWindowInternal$>$, IntHashMap$<$Integer$>$ (containing int and float)))), InputStream (containing int, byte[]), V, E, Url (containing List$<$String$>$, long, Integer, String, List$<$QueryParameter$>$ (QueryParameter contains String)), Serializable \\
\hline 
\emph{Apache storm 1.2.2} &  ObjectInputStream, ClassLoaderObjectInputStream, KerberosTicket (containing int, byte[], boolean[], Date, KeyImpl (containing transient variables => not considered), KerberosPrincipal (containing int and char), InetAddress[] (containing int, List$<$NameService$>$)), Kryo (but this class does not implement java.io.Serializable) \\ 
\hline
\emph{Apache-ofbiz-16.11.05} &  ObjectInputStream, String, ByteArrayInputStream (containing byte and int), Thread \\
\hline
\emph{Atlassian-bamboo-5.9.7} &  ObjectInputStream, E, Hashing$<$K$>$, Equiv$<$K$>$, K, V, ClassTag (which is an interface) and A \\
\hline
\emph{Apache-xmlrpc-3.1.3}  &  Throwable (containing String, StackTraceElement[], StackTraceElement, List$<$Throwable$>$,  and Throwable), ObjectInputStream \\
\hline
\emph{Gradle-enterprise-maven-extension-1.5.3}  &   AuthScheme (which is an interface), TestListenerEvent (containing long and TestDescriptor (with Long and String))      \\
\hline
\emph{Apache Chainsaw}  &    LoggingEvent (containing long, String, Hashtable(with int, float, sun.misc.Unsafe), boolean, ThrowableInformation (containing String[]), LocationInfo (containing String, Method, StringWriter (containing StringBuffer), PrintWriter (containing Writer (containing char[], int, Object), boolean, Formatter (containing Appendable (chich is an interface), Locale (containing Cache, char, int and Locale), IOException, char, double, int), PrintStream (containing boolean, Formatter, BufferedWriter (containing Writer (containing char[], int, Object), char, int, String ), OutputStreamWriter (containing StreamEncoder) ), String), boolean,Map$<$K,V$>$,  LogPanelPreferenceModel (containing String, Collection, ArrayList, boolean), ObjectInputStream, Point (containing int), Dimension (containing int), Vector (Object[], int)   \\
\hline
\emph{Jackson-databind-2.9.10.6}  &  ObjectIdReader (containing JavaType (which an abstract class contaning Class$<$?$>$, Object, int and boolean), PropertyName (containing long, String, PropertyName, SerializableString (which is an interface)), ObjectIdGenerator$<$?$>$, ObjectIdResolver (which is an interface), JsonDeserializer$<$Object$>$ (which is an abstract class), SettableBeanProperty (containing JsonDeserializer$<$Object$>$, PropertyName, JavaType, JsonDeserializer$<$Object$>$, TypeDeserializer (which is an abstract class), NullValueProvider (which is an interface))   \\
\hline
\emph{Opennms-source-26.0.1-1}  &   TrapInformation, T, OnmsCriteria, SerializedBatch, OnmsSeverity (containing Map$<$Integer, OnmsSeverity$>$, int, String)  \\
\hline
\emph{TeamCity-2019.1.3}  &   ObjectInputStream, Serializable, Principal (which is an interface), String[]  \\
\hline
\caption{Serialized and deserialized types extracted from vulnerable real-world applications} 
\label{tab:apps-type-extract}
\end{longtable}
\end{landscape}

\begin{landscape}
\section{Analysis of commits} 
\label{analyze-commits}

\footnotesize {
\begin{longtable}{|p{0.15\textwidth}|p{0.15\textwidth}|p{0.3\textwidth}|p{0.35\textwidth}|p{0.35\textwidth}|}
\hline
\textbf{Application} & \textbf{CVE} & \textbf{Open source? (Y/N)} &   \textbf{Internal/External patch?} & \textbf{Automatically/manually generated patch?} \\
\hline
Taoensso Nippy $<$  2.14.2 & CVE-2020-24164 & Y \url{https://github.com/ptaoussanis/nippy} &  Internal patch  & Manually generated patch (see \href{https://github.com/ptaoussanis/nippy/commit/61fb009fdde2994140f2da2e495ba8af3a873eb2}{Nippy commit} and \url{http://ptaoussanis.github.io/nippy/taoensso.nippy.html#var-allow-and-record-any-serializable-class-unsafe})  \\
\hline
Jboss RichFaces (Jboss-RF) 3.x $<=$ 3.3.3 and 4.x $<=$ 4.3.2 & CVE-2013-2165 & Y \url{https://richfaces.jboss.org/download/archive.html} & Internal patch  \url{https://www.bleathem.ca/blog/richfaces-security-advisory-cve-2013-2165/} and \url{https://codewhitesec.blogspot.com/2018/05/poor-richfaces.html}  & Manually generated patch (see \href{https://github.com/richfaces4/core/commit/12ee1166f04806b3ba072d27f9a9b3b3feae2ec9}{Jboss Richfaces commit}) \\
\hline
Android $<$ 5.0.0 & CVE-2014-7911 & Y \href{https://android.googlesource.com/?format=HTML}{https://android.googlesour\-ce.com/?format=HTML} & Internal patch  (see   \href{https://android.googlesource.com/platform/libcore/+/738c833d38d41f8f76eb7e77ab39add82b1ae1e2}{android libcore commit}) & Manually generated patch (see  \href{https://android-review.googlesource.com/c/platform/libcore/+/101525}{Android commit})  \\
\hline
Atlassian Bamboo before 5.9.9 and 5.10.x before 5.10.0 & CVE-2014-9757 & Y \url{https://www.atlassian.com/software/bamboo/download-archives} & External (the patched version Bamboo 5.10.0 uses an updated version of the smack library in which a lot of modifications are brought) & - \\
\hline
Jenkins $<$ 1.638 and LTS $<$ 1.625.2 & CVE-2015-8103 & Y \url{https://github.com/jenkinsci/jenkins} \url{https://wiki.jenkins.io/display/JENKINS/Jenkins+CLI} &  Unknown patch  & -  \\
\hline
VMware vRealize Orchestrator 6.x, vCenter Orchestrator 5.x, vRealize Operations 6.x, vCenter Operations 5.x, and vCenter Application Discovery Manager (vADM) 7.x & CVE-2015-6934 & Y \url{https://docs.vmware.com/en/vRealize-Orchestrator/7.6/rn/VMware-vRealize-Orchestrator-76-Release-Notes.html} & External patch (replace the Commons collections library by the commons-collections-3.2.2.jar in the dependencies of the mentioned products (see \url{https://kb.vmware.com/s/article/2141244} and \url{https://kb.vmware.com/s/article/2141244}) & -   \\
\hline
Hazelcast $<$ 3.11 & CVE-2016-10750 & Y \url{https://github.com/hazelcast/hazelcast} & Internal patch (see \url{https://docs.hazelcast.org/docs/3.10.5/manual/html-single/index.html#untrusted-deserialization-protection}) & Unknown  \\
\hline
Apache OFBiz 12.04.x $<$ 12.04.06 and 13.07.x $<$ 13.07.03 & CVE-2016-2170 & Y \url{http://archive.apache.org/dist/ofbiz/} & External patch (Update commons collections to 4.1 and Comment out RMI related code (see \url{https://issues.apache.org/jira/browse/OFBIZ-6942},  \url{https://markmail.org/message/nh6csf4fun5n6e23} and \url{https://issues.apache.org/jira/browse/OFBIZ-6726}) & - \\
\hline
Apache XML-RPC (aka ws-xmlrpc) library 3.1.3 & CVE-2016-5003 & Y \url{https://archive.apache.org/dist/ws/xmlrpc/sources/} & Internal patch & Manually generated patch  (see \href{https://src.fedoraproject.org/rpms/xmlrpc/c/ef4efbf91d241070f6f41950f7536049688a3a67?branch=master}{XML-RPC commit}) \\
\hline
Apache Wicket 6.x $<$ 6.25.0 and 1.5.x $<$ 1.5.17 & CVE-2016-6793 & Y \url{https://archive.apache.org/dist/wicket/} & Internal patch (obtained by doing the diff betwwen the 6.24.0 and 6.25.0 versions) & Manually generated patch (see \href{https://github.com/apache/wicket/commit/134686ef7185d3f96fec953136ab4847cd36b68}{wicket commit} \href{https://github.com/apache/wicket-site/commit/c202a1f616f460643bf82441480946e3f689f884}{Announcing CVE-2016-6793: Apache Wicket deserialization vulnerability})  \\
\hline
Red Hat JBoss Enterprise Application Platform (Jboss-EAP) 4 and 5 & CVE-2016-7065 & Y \url{https://developers.redhat.com/products/eap/download}  & Red Hat does not fix the issue because JBoss EAP 4 is out of maintenance
support and JBoss EAP 5 is close to the end of its maintenance period (see \url{https://seclists.org/fulldisclosure/2016/Nov/143} and \url{https://seclists.org/fulldisclosure/2016/Nov/143}) & - \\
\hline
Log4j  & CVE-2019-17571 & Y \url{https://github.com/apache/log4j} &  Internal & Manually generated patch (see \href{https://github.com/apache/logging-log4j2/commit/5dcc19215827db29c993d0305ee2b0d8dd05939d}{log4j commit})  \\
\hline
JetBrains TeamCity before 2019.1.4 & CVE-2019-18364 & Y \url{https://www.jetbrains.com/fr-fr/teamcity/download/other.html} & Unknown patch  & - \\
\hline
Apache Dubbo 2.7.0 before 2.7.5, 2.6.0 before 2.6.8, and 2.5.x versions & CVE-2019-17564 & Y \url{https://github.com/apache/dubbo} &  Internal & Manually generated patch (see \href{https://github.com/apache/dubbo/commit/9b18fe228971eaeca9b87d7b7e95df1c2a8ff91b}{Dubbo commit} and \url{https://github.com/apache/dubbo/releases/tag/dubbo-2.7.5})  \\
\hline
WebSphere Application Server (WAS) Community Edition 3.0.0.3 & CVE-2013-1777 & Y \url{http://geronimo.apache.org/downloads.html} & Internal patch  \url{http://svn.apache.org/viewvc?view=revision&sortby=date&revision=1458113} & Manually generated patch (\url{http://svn.apache.org/viewvc?view=revision&sortby=date&revision=1458113}) \\
\hline
 Apache Ofbiz from 16.11.01 to 16.11.05  & CVE-2019-0189 & Y \url{https://archive.apache.org/dist/ofbiz/} & Internal & Manually generated patch (see \href{https://github.com/apache/ofbiz-framework/commit/40c971275a743bcbad6a5384fd2c9cbfd6e80239}{Ofbiz commit}) \\
\hline
Apache Tomcat & CVE-2020-9484 & Y \url{https://github.com/apache/tomcat} & Internal & Manually generated patch (see \href{https://github.com/apache/tomcat/commit/bb33048e3f9b4f2b70e4da2e6c4e34ca89023b1b#diff-d2801d6b9c9ff6f98a6871accb7e61499ed3899f5234028997387ad65906e5e7}{Tomcat commit}\\
\hline
OpenNMS Horizon $<$ 26.0.1 and Meridian before 2018.1.19 and 2019 before 2019.1.7 & CVE-2020-12760 & Y \url{https://github.com/OpenNMS/opennms/releases/tag/opennms-26.0.1-1} &  External patch (see    \url{https://github.com/OpenNMS/opennms/pull/2983}) & - \\
\hline
Apache Tapestry 4 & CVE-2020-17531 & Y \url{https://github.com/apache/tapestry4}  & No available patch (Apache Tapestry 4 reached end of life in 2008 and no update to address this issue is released (the upgrade to the latest Apache Tapestry 5 version is necessary) (see \href{https://lists.apache.org/thread/mcl3xzw50vjb7rv76nsgq5zorhbg5gyy}{Tapestry 4 message}) and \url{https://cve.mitre.org/cgi-bin/cvename.cgi?name=CVE-2020-17531}) & -  \\
\hline
Apache Camel Netty (Camel-Netty) & CVE-2020-11973 & Y \url{https://github.com/apache/camel/tree/main/components/camel-netty} & Internal & Manually generated patch (see \href{https://github.com/apache/camel/pull/3537/commits/58aff9f9cd4aab3163b8eda8281cb795cb3b59c8}{Camel Netty commit})  \\
\hline
Apache Camel RabbitMQ (Camel-RabbitMQ) 2.22.x, 2.23.x, 2.24.x, 2.25.0, 3.0.0 up to 3.1.0  & CVE-2020-11972 & Y \url{https://github.com/apache/camel} & Internal patch (see \href{https://github.com/zregvart/camel/commit/c15ed20d92b5c920e9e55fe584f8e412b23f14f6}{RabbitMQ patch}) & Manually generated patch (see \href{https://github.com/apache/camel/pull/3633}{Camel commit})  \\
\hline
Emissary 6.4.0 & CVE-2021-32634 & Y \url{https://github.com/NationalSecurityAgency/emissary} &  Internal patch (\url{https://github.com/NationalSecurityAgency/emissary/security/advisories/GHSA-m5qf-gfmp-7638}) & Manually generated patch (see \href{https://github.com/NationalSecurityAgency/emissary/commit/40260b1ec1f76cc92361702cc14fa1e4388e19d7}{emissary commit}) \\
\hline
Apache Dubbo prior to 2.6.9 and 2.7.9 & CVE-2021-30179 & Y \url{https://github.com/apache/dubbo} & Internal patch (see \url{https://github.com/apache/dubbo/releases/tag/dubbo-2.7.10}) & Manually generated patch (see \href{https://github.com/apache/dubbo/pull/7436/commits/390c80a502dc7c2c6f29d544639760bcd2dc54fb}{dubbo commit} and \url{https://github.com/apache/dubbo/pull/7436})\\
\hline
Soffid IAM $<$ 1.7.5 & CVE-2017-9363 & Y \url{https://github.com/SoffidIAM/console} &  Internal patch & Manually generated patch (see \href{https://github.com/SoffidIAM/console/commit/8e9e7c9e537acfc2a245fbbeb41a143b5b4f7230#diff-544c1cb1ac64f2f62b6b326bd0b1b6addc17f19416878d319d3643e302a043b7}{SoffidIam commit})   \\
\hline
\multirow{2}{*}{Apache OFBiz} & CVE-2021-29200  & Y \url{https://github.com/apache/ofbiz-framework} & Internal & Manually generated patch (see \href{https://github.com/apache/ofbiz-framework/commit/1bc8a206346f251c9076c2fb9babc896ba6bdf0d}{Ofbiz commit})  \\
\cline{2-5}
& CVE-2021-26295 & Y \url{https://github.com/apache/ofbiz-framework} & Internal & Manually generated patch (see \url{https://issues.apache.org/jira/browse/OFBIZ-12167} and \href{https://github.com/apache/ofbiz-plugins/commit/a3438121d8f50545b3a5c397c589fe97ca33202b}{Ofbiz commit})  \\
\hline
\caption{Analysis of the patches of 25 open source vulnerable applications. In this table, the fourth column "Internal/External patch?" describes if the patch concerns the code of the application itself (Internal) or the code of the libraries used in the concerned application (External). The fifth column  "Automatically/manually generated patch?" indicates if the patch was manually generated or automatically generated by tools like Snyk~\citep{snyk,snykGithub}. For each existing patch,  we give the link to the commit in which is described the patch. For the applications for which we do not find commit for patch, we use the keyword "Unknown" in the last column. For the applications that do not have a patch, we put the "-" symbol.}
  \label{tab:patches-commits}
\end{longtable}
}
\end{landscape}

\end{appendices}

\end{document}